\documentclass[pra,showpacs,nofootinbib]{revtex4}
\usepackage{graphicx}

\def\be{\begin{equation}}
\def\ee{\end{equation}}
\def\ba{\begin{eqnarray}}
\def\ea{\end{eqnarray}}

\begin{document}
\title{Density-functional-theory calculations of matter in strong
magnetic fields. II. Infinite chains and condensed matter}
\author{Zach Medin and Dong Lai}
\affiliation{Center for Radiophysics and Space Research, Department of
Astronomy, Cornell University, Ithaca, New York 14853, USA}

\received{12 July 2006; published 14 December 2006}

\begin{abstract}
We present calculations of the electronic structure of one-dimensional
infinite chains and three-dimensional condensed matter in strong
magnetic fields ranging from $B=10^{12}$~G to $2\times10^{15}$~G,
appropriate for observed magnetic neutron stars.  At these field
strengths, the magnetic forces on the electrons dominate over the
Coulomb forces, and to a good approximation the electrons are confined
to the ground Landau level. Our calculations are based on the density
functional theory, and use a local magnetic exchange-correlation
function appropriate in the strong field regime.  The band structures
of electrons in different Landau orbitals are computed
self-consistently.  Numerical results of the ground-state energies and
electron work functions are given for one-dimensional chains
H$_\infty$, He$_\infty$, C$_\infty$, and Fe$_\infty$. Fitting formulae
for the $B$-dependence of the energies are also provided.  For all the
field strengths considered in this paper, hydrogen, helium, and carbon
chains are found to be bound relative to individual atoms (although
for $B$ less than a few $\times 10^{12}$~G, carbon infinite chains are
very weakly bound relative to individual atoms). Iron chains are
significantly bound for $B\agt 10^{14}$~G and are weakly bound if at
all at $B\alt 10^{13}$~G\@. We also study the cohesive property of
three-dimensional condensed matter of H, He, C, and Fe at zero
pressure, constructed from interacting chains in a body-centered
tetragonal lattice. Such three-dimensional condensed matter is found
to be bound relative to individual atoms, with the cohesive energy
increasing rapidly with increasing $B$.
\end{abstract}


\maketitle

%%%%%%%%%%%%%%%%%%%%%%%%%%%%%%%%%%%%%%%%%%%%%%%%%%%%
\section{Introduction}
\label{sec:intro}

Young neutron stars (ages $\alt 10^7$~years) are observed to 
have surface magnetic fields in the range of $10^{11}$-$10^{15}$~G
\citep{meszaros92,reisenegger05,woods05,harding06}, far beyond the reach
of terrestrial laboratories \citep{wagner04}.
It is well known that the properties of matter can be drastically
modified by such strong magnetic fields. The
natural atomic unit for the magnetic field strength, $B_0$, is set by
equating the electron cyclotron energy $\hbar\omega_{Be}=\hbar
(eB/m_ec)=11.577\,B_{12}$~keV, where $B_{12}=B/(10^{12}~{\rm G})$, to
the characteristic atomic energy $e^2/a_0=2\times 13.6$~eV (where
$a_0$ is the Bohr radius): 
\be
B_0={m_e^2e^3c\over\hbar^3}=2.3505\times 10^9\, {\rm G}.
\label{eqb0}
\ee
For $b=B/B_0\agt 1$, the usual perturbative treatment of the magnetic
effects on matter (e.g., Zeeman splitting of atomic energy levels)
does not apply. Instead, in the transverse direction (perpendicular to
the field) the Coulomb forces act as a perturbation to the magnetic
forces, and the electrons in an atom settle into the ground Landau
level. Because of the extreme confinement of the electrons in the
transverse direction, the Coulomb force becomes much more effective in
binding the electrons along the magnetic field direction. The atom
attains a cylindrical structure. Moreover, it is possible for these
elongated atoms to form molecular chains by covalent bonding along the
field direction. Interactions between the linear chains can then lead
to the formation of three-dimensional condensed matter
\citep{ruderman74,ruder94,lai01}.

This paper is the second in a series where we present calculations of
matter in strong magnetic fields using density functional theory. In
\citet{medin06a} (hereafter paper I), we studied various atoms and
molecules in magnetic fields ranging from $10^{12}$~G to $2\times
10^{15}$~G for H, He, C, and Fe, representative of the most likely
neutron star surface compositions. Numerical results and fitting
formulae of the ground-state energies were given for H$_N$ (up to
$N=10$), He$_N$ (up to $N=8$), C$_N$ (up to $N=5$), and Fe$_N$ (up to
$N=3$), as well as for various ionized atoms. It was found that as $B$
increases, molecules become increasingly more bound relative to
individual atoms, and that the binding energy per atom in a molecule,
$|E_N|/N$, generally increases and approaches a constant value with
increasing $N$.  In this paper, we present density-functional-theory
calculations of infinite chains of H, He, C, and Fe. Our goal is to
obtain the cohesive energy of such one-dimensional (1D) condensed
matter relative to individual atoms for a wide range of field
strengths. We also carry out approximate calculations of the relative
binding energy between 1D chains and three-dimensional (3D) condensed
matter at zero pressure.

The cohesive property of matter in strong magnetic fields is a
fundamental quantity characterizing magnetized neutron star surface
layers, which play a key role in many neutron star processes and
observed phenomena. The cohesive energy refers to the energy required
to pull an atom out of the bulk condensed matter at zero pressure.
Theoretical models of pulsar and magnetar magnetospheres depend on the
cohesive properties of the surface matter in strong magnetic fields
\citep{ruderman75,arons79,usov96,harding98,beloborodov06,gil03}.  For
example, depending on the cohesive energy of the surface matter, an
acceleration zone (``polar gap'') above the polar cap of a pulsar may
or may not form, and this will affect pulsar radio emission and other
high-energy emission processes. Also, while a hot or warm neutron star most
certainly has a gaseous atmosphere that mediates its thermal emission,
condensation of the stellar surface may occur at sufficiently low
temperatures \citep{lai97,lai01}. For example, radiation from a bare
condensed surface (with no atmosphere above it) has been invoked to
explain the nearly perfect blackbody emission spectra observed in some
nearby isolated neutron stars
\citep{burwitz03,mori03,vanadelsberg05,turolla04,perez06}.  However,
whether surface condensation actually occurs depends on the cohesive
energy of the surface matter.

There have been few quantitative studies of infinite chains and
zero-pressure condensed matter in strong magnetic fields.  Earlier
variational calculations \citep{flowers77,muller84} as well as
calculations based on Thomas-Fermi type statistical models
\citep{abrahams91,fushiki92,lieb94a,lieb94b}, while useful in
establishing scaling relations and providing approximate energies of
the atoms and the condensed matter, are not adequate for obtaining
reliable energy differences (cohesive energies).  Quantitative results
for the energies of infinite chains of hydrogen molecules H$_\infty$
in a wide range of field strengths ($B\gg B_0$) were presented in both
Ref.~\citep{lai92} (using the Hartree-Fock method with the plane-wave
approximation; see also Ref.~\citep{lai01} for some results of
He$_\infty$) and Ref.~\citep{relovsky96} (using density functional
theory). For heavier elements such as C and Fe, the cohesive energies
of 1D chains have only been calculated at a few magnetic field
strengths in the range of $B=10^{12}$-$10^{13}$~G, using Hartree-Fock
models \citep{neuhauser87} and density functional theory
\citep{jones85}. There were discrepancies between the results of these
works, and some (e.g., Ref.~\citep{neuhauser87}) adopted a crude
treatment for the band structure (see Sec.~\ref{subsec:complex}). An
approximate calculation of 3D condensed matter based on density
functional theory was presented in Ref.~\citep{jones86}.

Our calculations of atoms and small molecules (paper I) and of
infinite chains and condensed matter (this paper) are based on density
functional theory \citep{hohen64,kohn65, vignale87,vignale88,jones89}.
In the strong field regime where the electron spins are aligned with
each other, the Hartree-Fock method is expected to be highly accurate
\citep{neuhauser87,schmelcher99}. However, in dealing with systems
with many electrons, it becomes increasingly impractical as the
magnetic field increases, since more and more Landau orbitals (even
though electrons remain in the ground Landau level) are occupied and
keeping track of the direct and exchange interactions between
electrons in various orbitals becomes computational rather tedious.
Our density-functional calculations allow us to obtain the energies of
atoms and small molecules and the energy of condensed matter using the
same method, thus providing reliable cohesive energy values for
condensed surfaces of magnetic neutron stars, a main goal of our
study. Compared to previous density-functional theory calculations
\citep{jones85,jones86, kossl88,relovsky96}, we use an improved
exchange-correlation function appropriate for highly magnetized
electron gases, we calibrate our density-functional code with previous
results (when available) based on other methods, and (for calculations
of condensed matter) adopt a more accurate treatment of the band
structure. Moreover, our calculations extend to the magnetar-like
field regime ($B\sim 10^{15}$~G).

This paper is organized as follows.  After briefly summarizing the
approximate scaling relations for linear chains and condensed matter
in strong magnetic fields in Sec.~II, we describe our method and the
basic equations in Sec.~III\@.  Numerical results (tables and fitting
formulae) for linear chains are presented in Sec.~IV\@. In Sec.~V we
describe our approximate calculation and results for the relative
energy between 1D chain and 3D condensed matter.  We conclude in
Sec.~VI\@. Some technical details are given in the appendix.

%%%%%%%%%%%%%%%%%%%%%%%%%%%%%%%%%%%%%%%%%%%%%%%%%%%%%%%%%%%%%%%%%%%%%%%
\section{Basic scaling relations for linear chains and 3D condensed matter 
in strong magnetic fields}
\label{sec:basic}

The simplest model for the linear chain is to treat it as a uniform
cylinder of electrons, with ions aligned along the magnetic field
axis. The radius of the cylinder is $R$ and the length of a unit cell
is $a$ (which is also the atomic spacing along the $z$ axis).  The
electrons lie in the ground Landau level, but can occupy different
Landau orbitals with the radius of guiding center
$\rho_m=(2m+1)^{1/2}\rho_0$, where $m=0,1,2,\ldots,m_{\rm max}$ and
$\rho_0=(\hbar c/eB)^{1/2}=b^{-1/2}$ (in atomic
units).\footnote{Unless otherwise specified, we use atomic units, in
which the length in $a_0$ (Bohr radius), mass in $m_e$, energy in
$e^2/a_0=2$~Ry, and magnetic field strength in units of $B_0$.} The
maximum Landau orbital number $m_{\rm max}$ is set by $\rho_{m_{\rm
max}}=R$, giving $m_{\rm max}\simeq \pi R^2 eB/(hc)=R^2 b/2$ (this
is the Landau degeneracy in area $\pi R^2$). For a uniform electron
density $n=Z/(\pi R^2 a)$, the Fermi wave number (along $z$) $k_F$ is
determined from $n=bk_F/(2\pi^2)$, and the kinetic energy of the
electrons in a cell is $E_k=(Z/3)\varepsilon_F'$, with
$\varepsilon_F'=k_F^2/2$ the Fermi kinetic energy.  The total energy
per atom (unit cell) in the chain can be written as
\citep{ruderman71,ruderman74}
\be
E_\infty={2\pi^2Z^3\over 3b^2R^4a^2}-{Z^2\over a}
\left[\ln{2a\over R}-\left(\gamma-{5\over 8}\right)\right],
\label{einft}
\ee
where $\gamma=0.5772\ldots$ is Euler's constant.
In Eq.~(\ref{einft}), the first term is the electron kinetic energy $E_k$
and the second term is the (direct) Coulomb energy 
(the Madelung energy for the one-dimensional uniform lattice). 
Minimizing $E_\infty$ with respect to $R$ and $a$ gives
\ba
&&R=1.65\,Z^{1/5}b^{-2/5},\qquad a/R=2.14,\nonumber\\
&&E_\infty=-0.354\,Z^{9/5}b^{2/5}.
\label{chainscale}\ea
Note that the energy (\ref{einft}) can be written as
$E_\infty=-ZV_0+(Z/3)\varepsilon_F'$, where $V_0$ is the depth of the
potential well relative to the continuum. In equilibrium
$E_\infty=-5E_k=-(5/3)Z\varepsilon_F'$, and thus
$V_0=2\varepsilon_F'$. The Fermi level energy of the electrons in the
chain relative to the continuum is then
$\varepsilon_F=\varepsilon_F'-V_0=-\varepsilon_F'=3E_\infty/(5Z)$,
i.e.,
\be
\varepsilon_F(1D) = -0.212\,Z^{4/5}b^{2/5} \mbox{ a.u.} 
= -65.1\,Z^{4/5}B_{12}^{2/5} \mbox{ eV.}
\label{fermiscale}
\ee
Alternatively, if we identify the number of electrons in a cell,
$N_e$, as an independent variable, we find
$R=1.65\,(N_e^2/Z)^{1/5}b^{-2/5}$ and
$E_\infty=-0.354\,(Z^2N_e)^{3/5}b^{2/5}$.  The chemical potential
(which includes potential energy) of electrons in the chain is simply
$\mu=\varepsilon_F=\partial E_\infty/\partial N_e$, in agreement with
Eq.~(\ref{fermiscale}). The electron work function is
$W=|\varepsilon_F|$.

A linear 1D chain naturally attracts neighboring chains through the
quadrupole-quadrupole interaction. By placing parallel chains close together
(with spacing of order $b^{-2/5}$), we obtain three-dimensional 
condensed matter (e.g., a body-centered tetragonal lattice) 
\citep{ruderman71}. 

The binding energy of the 3D condensed matter at zero pressure can be
estimated using the uniform electron gas model.  Consider a
Wigner-Seitz cell with radius $r_i=Z^{1/3}r_s$ ($r_s$ is the mean
electron spacing); the mean number density of electrons is
$n=Z/(4\pi r_i^3/3)$.
When the Fermi energy $p_F^2/(2m_e)$ is less than the electron
cyclotron energy $\hbar\omega_{Be}$, or when the electron number 
density satisfies
$n \le n_{B}=(\sqrt{2}\pi^2\rho_0^3)^{-1}=0.0716\,b^{3/2}$
(or $r_i\ge r_{iB}=1.49\,Z^{1/3}b^{-1/2}$), the electrons only occupy
the ground Landau level. The energy per cell can be written
\be
E_s(r_i)={3\pi^2Z^3\over 8b^2r_i^6}-{0.9Z^2\over r_i},
\label{estry}\ee
where the first term is the kinetic energy and the second term is the
Coulomb energy. For a zero-pressure condensed matter, we require
$dE_s/dr_i=0$, and the equilibrium $r_i$ and energy are then given by 
\ba
&&r_{i}\simeq 1.90\,Z^{1/5}b^{-2/5},\label{eqri0}\\
&&E_{s}\simeq -0.395\,Z^{9/5}b^{2/5}.\label{es0}
\ea
The corresponding zero-pressure condensation density is 
\be
\rho_{s}\simeq 561\,A\,Z^{-3/5}B_{12}^{6/5}\,{\rm g~cm}^{-3}.
\label{rs0}\ee
The electron Fermi level energy is
\be
\varepsilon_F(3D) = \frac{3}{5Z}E_s = -0.237\,Z^{4/5}b^{2/5} \mbox{ a.u.} 
= -72.7\,Z^{4/5}B_{12}^{2/5} \mbox{ eV.}
\label{fermiscale3D}
\ee
The uniform electron gas model can be improved by incorporating
the Coulomb exchange energy and Thomas-Fermi correction
due to nonuniformity of the electron gas \citep{lai01,fushiki89}.

Although the simple uniform electron gas model and its Thomas-Fermi
type extensions give a reasonable estimate for the binding energy for
the condensed state, they are not adequate for determining the
cohesive property of the condensed matter. Also, as we shall see,
Eq.~(\ref{fermiscale}) or Eq.~(\ref{fermiscale3D}) does not give a
good scaling relation for the electron work function when detailed
electron energy levels (bands) in the condensed matter are taken into
account. The cohesive energy $Q_s=E_a-E_s$ is the difference between
the atomic ground-state energy $E_a$ and the condensed matter energy
per cell $E_s$. In principle, a three-dimensional electronic band
structure calculation is needed to solve this problem. However, for
sufficiently strong magnetic fields, such that
$a_0/Z\gg\sqrt{2Z+1}\rho_0$ or $B_{12}\gg 100\,(Z/26)^3$, a linear 1D
chain is expected to be strongly bound relative to individual atoms
(i.e., the cohesive energy of the chain, $Q_\infty=E_a-E_\infty$, is
significantly positive) \citep{lai01}. For such strong fields, the
binding of 3D condensed matter results mainly from the covalent bond
along the magnetic axis, rather than from chain-chain interactions; in
another word, the energy difference $|\Delta E_s|=|E_s-E_\infty|$ is
small compared to $Q_\infty$. In the magnetic field regime where
$Q_\infty$ is small or even negative, chain-chain interactions are
important in deciding whether 3D condensed matter is bound relative to
individual atoms. In this paper we will concentrate on calculating
$E_\infty$ and $Q_\infty$ for linear chains (Sec.~III and Sec.~IV).
In Sec.~V we shall quantify the magnitude of $\Delta E_s$ for
different elements and field strengths.

%%%%%%%%%%%%%%%%%%%%%%%%%%%%%%%%%%%%%%%%%%%%%%%%%%%%%%%%%%%%%%%%%%%%%%%
\section{Density-functional-theory calculations of 1D chains: 
Methods and equations}
\label{sec:method}

Our calculations of 1D infinite chains are based on density functional
theory, which is well established in the strong magnetic field regime
($B\gg B_0$) of interest here \citep{vignale87,vignale88}. Extensive
comparisons of our density-functional-theory results for atoms and
finite molecules with previous results (when available) based on
different methods were given in paper I \citep{medin06a}; such
comparisons established the validity and calibrate the systematic
error of our approach. As we discuss below, for infinite chains
considered in the present paper, it is important to calculate the band
structure of electrons (for different Landau orbitals)
self-consistently, rather than using certain approximate 
{\it ans\"atze} as adopted in some previous works \citep{neuhauser87}.

%%%%%%%%%%%%%%%%%%%%%%%%%%%%%%%%%%%%%%%%%%%%%%%%%%
\subsection{Basic equations and concepts}
\label{subsec:eqn}

Our calculations will be based on the ``adiabatic approximation,'' 
in which all electrons are assumed to lie in the ground Landau level.
For elements with nuclear charge number $Z$, this is 
an excellent approximation for $b\gg Z^2$. Even under the more relaxed 
condition $b\gg Z^{4/3}$, this approximation is expected to yield a
reasonable total energy and accurate results for the 
energy difference between different electronic systems (atoms and chains)
(see paper I). Also, we use nonrelativisitc quantum mechanics in our
calculations, even when $\hbar\omega_{Be}\agt m_ec^2$ or $B\agt
B_Q=B_0/\alpha^2=4.414\times 10^{13}$~G\@. As discussed in paper I,
this is accurate as long as the electrons stay in the ground
Landau level.

In a 1D chain, the ions form a periodic lattice along the magnetic
field axis. The number of cells (``atoms'') in the chain is
$N\rightarrow\infty$ and the ions are equally spaced with lattice
spacing $a$.  In the adiabatic approximation, the one-electron
wave function (``orbital'') can be separated into a transverse
(perpendicular to the external magnetic field) component and a
longitudinal (along the magnetic field) component:
\be
\Psi_{m\nu k}(\mathbf{r}) = 
\frac{1}{\sqrt{N}}W_m(\mathbf{r_\perp})f_{m\nu k}(z)\,.  
\ee
Here $W_m$ is the ground-state Landau wave function \citep{landau77} given
by
\be W_m(\mathbf{r_\perp}) =
\frac{1}{\rho_0 \sqrt{2\pi m!}}
\left(\frac{\rho}{\sqrt{2}\rho_0}\right)^m
\exp\left(\frac{-\rho^2}{4\rho_0^2}\right) \exp(-im\phi) \,,
\label{Wmeq}
\ee
which is normalized as $\int\!d^2{\bf r}_\perp |W_m|^2=1$.
The longitudinal wave function $f_{m\nu k}$  must be solved
numerically, and we choose to normalize it over a unit cell of the lattice:
\be
\int_{-a/2}^{a/2} |f_{m\nu k}(z)|^2 \, dz = 1,
\ee
so that normalization of $\Psi_{m\nu k}$ is $\int\!d^3{\bf r}\,
|\Psi_{m\nu k}|^2=1$ (here and henceforth, the general
integral sign $\int\!d^3{\bf r}$ refers to integration over the whole
chain, with $z$ from $-Na/2$ to $Na/2$).
The index $\nu=0,1,2,\ldots$ labels the different bands of the electron
(see below), rather than the number of nodes in the longitudinal 
wave function as in the atom or molecule case. 

The quantum number $k$ is not present for atoms or finite molecules, but enters
here because of the periodic nature of the electrons
in the longitudinal direction. By Bloch's theorem, the electrons
satisfy the periodicity condition
\be
f_{m\nu k}(z+a) = e^{ika}f_{m\nu k}(z) \,,
\label{periodeq}
\ee
and $k$ is the Bloch wave number.
Note that the longitudinal wave functions are periodic in $k$ with
period $\Delta k = 2\pi/a$; i.e., $f_{m\nu,k+K}(z) = f_{m\nu k}(z)$
with $K$ being any reciprocal vector (number, in one dimension) of the
lattice, $K=2\pi n/a$ ($n$ is an integer).  Because of this, to ensure
that each wave function $f_{m\nu k}$ is unique, we restrict $k$ to the
first Brillouin zone, $k \in [-\pi/a,\pi/a]$.  The electrons fill each
$(m\nu)$ band, with spacing $\Delta k=\pi/(Na)$, and thus the maximum
number of electrons in a given band is $N$ (out of the total $ZN$
electrons in the chain). In another word, the number of electrons per
unit cell in each ($m\nu$) band is $\sigma_{m\nu}\le 1$ (see
Sec.~\ref{subsec:band}).

The density distribution of electrons in the chain is given by 
\be 
n(\mathbf{r}) = \sum_{m\nu k} |\Psi_{m\nu k}(\mathbf{r})|^2 =
\frac{a}{2\pi} \sum_{m\nu} |W_m|^2(\rho) \int_{I_{m\nu}} dk \,
|f_{m\nu k}(z)|^2 \,,
\label{eq:density}
\ee
where the sum/integral is over all electron states, each electron
occupying an $(m\nu k)$ orbital. The notation
$|W_m|^2(\rho)=|W_m(\mathbf{r}_\perp)|^2$ is used here because $W_m$
is a function of $\rho$ and $\phi$ but $|W_m|^2$ is a function of
$\rho$ only. The notation $\int_{I_{m\nu}}$ in the $k$ integral refers
to the fact that the region of integration depends on the $(m\nu)$
level; we will discuss this interval and electron occupations in
Sec.~\ref{subsec:band}.
To simplify the appearance of the electron density expression, 
we define the function 
\be
\bar{f}_{m\nu}(z) = \sqrt{\frac{a}{2\pi} \int_{I_{m\nu}} dk \,
|f_{m\nu k}(z)|^2} \,, 
\label{eq:density2}\ee 
so that \be n(\mathbf{r}) = \sum_{m\nu}
|W_m|^2(\rho) \bar{f}_{m\nu}^{\,2}(z) \,.
\label{densityeq}
\ee

In an external magnetic field, the Hamiltonian of a free electron is
\be
\hat{H} = \frac{1}{2m_e}\left(\mathbf{p}+\frac{e}{c}\mathbf{A}\right)^2+\frac{\hbar eB}{2m_e c}\sigma_z \,,
\ee
where $\mathbf{A}=\frac{1}{2}\mathbf{B}\times\mathbf{r}$ is the vector potential of the external magnetic field and $\sigma_z$ is the $z$-component Pauli spin matrix. For electrons in Landau levels, with their spins aligned parallel/antiparallel to the magnetic field, the Hamiltonian becomes
\be
\hat{H} = \frac{\hat{p}_z^2}{2m_e}+\left(n_L+\frac{1}{2}\right)\hbar\omega_{Be}\pm\frac{1}{2}\hbar\omega_{Be} \,,
\ee
where $n_L=0,1,2,\ldots$ is the Landau level index; for electrons in the ground Landau level, with their spins aligned antiparallel to the magnetic field (so $n_L=0$ and $\sigma_z \rightarrow -1$),
\be
\hat{H} = \frac{\hat{p}_z^2}{2m_e} \,.
\ee
The total Hamiltonian for the atom or molecule then becomes
\be
\hat{H} = \sum_i \frac{\hat{p}_{z,i}^2}{2m_e} + V \,,
\ee
where the sum is over all electrons and $V$ is the total potential energy of the atom or molecule.

In the density functional formalism, the total energy per cell of the chain
is expressed as a functional of the total electron density
$n(\mathbf{r})$:
\be
E[n] = E_K[n] + E_{eZ}[n] + E_{\rm dir}[n] + E_{\rm exc}[n] + E_{ZZ}[n] \,.
\ee
Here $E_K[n]$ is the kinetic energy of the system of non-interacting electrons,
and $E_{eZ}$, $E_{\rm dir}$ and $E_{ZZ}$ are the electron-ion Coulomb energy,
the direct electron-electron interaction 
energy and the ion-ion interaction energy, respectively:
\be
E_{eZ}[n] = -\!\!\sum_{j=-N/2}^{N/2} Ze^2 \int_{|z|<a/2} d\mathbf{r} \,
\frac{n(\mathbf{r})}{|\mathbf{r} - \mathbf{z}_j|} \,,
\ee
\be
E_{\rm dir}[n] = \frac{e^2}{2} \int \!\! \int_{|z|<a/2}
d\mathbf{r}\,d\mathbf{r}' \,
\frac{n(\mathbf{r}) n(\mathbf{r}')}{|\mathbf{r} - \mathbf{r}'|} \,,
\ee
\be
E_{ZZ}[n] = \sum_{j=1}^{N/2} \frac{Z^2e^2}{ja} \,.
\ee
The location of the ions in the above equations is represented by the set
$\{\mathbf{z}_j\}$, with
\be
{\bf z}_j = ja\hat{\mathbf{z}}, \quad j=(-N/2),(-N/2+1),\ldots,0,\ldots,N/2.
\ee
The term $E_{\rm exc}$ represents the exchange-correlation energy. 
In the local approximation, 
\be
E_{\rm exc}[n] = \int_{|z|<a/2} \! d\mathbf{r} \, n(\mathbf{r})\, 
\varepsilon_{\rm exc}(n) \,,
\ee 
where $\varepsilon_{\rm exc}(n)=\varepsilon_{\rm ex}(n)+\varepsilon_{\rm corr}(n)$
is the exchange and correlation energy
per electron in a uniform electron gas of density $n$. 
For electrons in the ground Landau level, the (Hartree-Fock)
exchange energy can be written as \citep{danz71}
\be 
\varepsilon_{\rm ex}(n) = -\pi e^2 \rho_0^2
n F(t) \,, 
\ee 
where the dimensionless function $F(t)$ is 
\be
F(t)=4\int_0^\infty\!\!\!dx\left[\tan^{-1}\left({1\over
x}\right)-{x\over 2}\ln\left(1+{1\over x^2}\right)\right] e^{-4tx^2},
\ee
and 
\be 
t =\left(\frac{n}{n_B}\right)^2 = 2\pi^4 \rho_0^6 n^2,
\label{eq:tdefine}\ee
[$n_B=(\sqrt{2}\pi^2\rho_0^3)^{-1}$ is the density above which 
the higher Landau levels start to be filled in a uniform electron gas].
For small $t$, $F(t)$ can be expanded as \citep{fushiki89} 
\be 
F(t) \simeq 3-\gamma-\ln 4t +
\frac{2t}{3}\left(\frac{13}{6}-\gamma-\ln 4t \right) +
\frac{8t^2}{15}\left(\frac{67}{30}-\gamma-\ln 4t \right)
+{\cal O}(t^3\ln t),
\ee
where $\gamma= 0.5772\ldots$ is Euler's constant. We have found that
the condition $t\ll 1$ is well satisfied everywhere for almost all
infinite chains in our calculations. The notable exceptions are the
carbon chains at $B=10^{12}$~G and the iron chains at
$B\le 10^{13}$~G, which have $t\alt 1$ near the center of each cell. These
chains are expected to have higher $t$ values than the other chains in
our calculations, as they have large $Z$ and low $B$ \footnote{For the
uniform gas model, $t \propto Z^{4/5}B^{-3/5}$.}.

The correlation energy of uniform electron gas in strong magnetic
fields has not be calculated in general, except in the regime $t\ll 1$
and Fermi wave number $k_F=2\pi^2\rho_0^2 n \gg 1$ [or $n\gg
(2\pi^3\rho_0^2 a_0)^{-1}$]. \citet{skud93} use the random-phase
approximation to find a numerical fit for the correlation energy in
this regime (see also Ref.~\citep{stein98}):
\be \varepsilon_{\rm corr} =
-\frac{e^2}{\rho_0}\,[0.595 (t/b)^{1/8} (1-1.009t^{1/8})] \,.
\label{correq}
\ee
In the absence of an ``exact'' correlation energy density we employ
this strong-field-limit expression. Fortunately, because we are
concerned mostly with finding the energy difference between 
atoms and chains, the correlation energy term does not have to
be exact. The presence or the form of the correlation term has a
modest effect on the atomic and chain energies calculated but has
very little effect on the energy difference between them (see paper I
for more details on various forms of the correlation energy and 
comparisons).

Variation of the total energy with respect to the electron
density, $\delta E[n]/\delta n=0$, leads to the Kohn-Sham equation:
\be 
\left[ -\frac{\hbar^2}{2m_e} \nabla^2 + V_{\rm eff}(\mathbf{r}) \right]
\Psi_{m\nu k}(\mathbf{r}) =
\varepsilon_{m\nu}(k) \Psi_{m\nu k}(\mathbf{r}) \,, 
\ee
where
\be
V_{\rm eff}(\mathbf{r}) =
-\!\!\sum_{j=-N/2}^{N/2} \frac{Ze^2}{|\mathbf{r}-\mathbf{z}_j|}
+ e^2 \int d\mathbf{r}' \, \frac{n(\mathbf{r}')}{|\mathbf{r} - \mathbf{r}'|}
+ \mu_{\rm exc}(n),
\ee
with
\be
\mu_{\rm exc}(n) = \frac{\partial (n \varepsilon_{\rm exc})}{\partial n} \,.
\ee
Averaging the Kohn-Sham equation over the transverse wave function yields
a set of one-dimensional equations:
\ba 
\left[ -\frac{\hbar^2}{2m_e}\frac{d^2}{dz^2}+
{\bar V}_{\rm eff}(z)\right]
f_{m\nu k}(z) = \varepsilon_{m\nu}(k) f_{m\nu k}(z) \,.
\label{kohneq}
\ea
where 
\ba 
&& {\bar V}_{\rm eff}(z)=
- Ze^2\!\sum_{j=-N/2}^{N/2} \int d\mathbf{r}_\perp \,
\frac{|W_m|^2(\rho)}{|\mathbf{r}-\mathbf{z}_j|}
+ e^2 \int \!\! \int d\mathbf{r}_\perp\,d\mathbf{r}' \,
\frac{|W_m|^2(\rho)\, n(\mathbf{r}')}{|\mathbf{r} - \mathbf{r}'|}\nonumber\\
&&\qquad \qquad
+ \int d\mathbf{r}_\perp \, |W_m|^2(\rho)\, \mu_{\rm exc}(n). 
\label{kohneq2}
\ea
This set of equations are solved self-consistently to find the eigenvalue
$\varepsilon_{m\nu}(k)$ and the longitudinal wave function
$f_{m\nu k}(z)$ for each orbital occupied by the electrons.
Once these are known, the total energy per cell of the infinite
chain can be calculated using
\ba
E_\infty & = & \frac{a}{2\pi} \sum_{m\nu} \int_{I_{m\nu}} dk \,
 \varepsilon_{m\nu}(k) - \frac{e^2}{2} \int \!\! \int_{|z|<a/2}
 d\mathbf{r} d\mathbf{r}' \, \frac{n(\mathbf{r})
 n(\mathbf{r}')}{|\mathbf{r} - \mathbf{r}'|} \nonumber\\
& & +
 \int_{|z|<a/2} d\mathbf{r} \, n(\mathbf{r}) [\varepsilon_{\rm
 exc}(n)-\mu_{\rm exc}(n)] + \sum_{j=1}^{N/2} \frac{Z^2e^2}{ja} \,,
\label{energyeq}
\ea
where the interval $I_{m\nu}$ is the same as in the electron 
density expression, Eq.~(\ref{densityeq}). 

Note that the electron-ion, direct electron-electron, and ion-ion
interaction energy terms given above formally diverge for
$N\rightarrow\infty$.  These terms must be properly combined to yield
a finite net potential energy. Note that for an electron in the
``primary'' unit cell ($-a/2\le z\le a/2$), the potential generated by
a distant cell (centered at $z_j=ja$) can be well approximated by the
quadrupole potential: 
\be 
V_{Q}(\rho,z;ja)=
\frac{3e^2}{2}\frac{Q_{zz}}{|ja|^5} \left( 2z^2-\rho^2 \right), 
\ee 
where $Q_{zz}$ is the quadrupole moment of a unit cell 
\be 
Q_{zz} = \int_{|z|<a/2} d\mathbf{r} \, 
\left( 2z^2-\rho^2 \right) n(\rho,z)
\,.  
\label{eq:qzz}\ee
The Coulomb (quadrupole-quadrupole) 
energy between the primary cell and the distant cell is simply 
\be
E_{QQ}(ja)=\int_{|z|<a/2} d\mathbf{r} \, 
n({\bf r})\,V_Q(\rho,z;ja)=
\frac{3e^2}{2}\frac{Q_{zz}^2}{|ja|^5} \,.
\ee
In our calculations, we treat distant cells with $|j|>N_Q$ using the
quadrupole approximation, while treating the nearby cells ($|j|\le N_Q$) 
exactly.
Thus the (averaged) effective potential, Eq.~(\ref{kohneq2}), becomes
\ba
&& {\bar V}_{\rm eff}(z)=
 -Ze^2\!\sum_{j=-N_Q}^{N_Q}\int d\mathbf{r}_\perp \,
 \frac{|W_m|^2(\rho)}{|\mathbf{r}-\mathbf{z}_j|} 
+ e^2 \int \!\! \int_{|z'|<a(N_Q+1/2)}
 d\mathbf{r}_\perp\,d\mathbf{r}' \, \frac{|W_m|^2(\rho)
 n(\mathbf{r}')}{|\mathbf{r} - \mathbf{r}'|} \nonumber\\
&& + \int d\mathbf{r}_\perp
 \, |W_m|^2(\rho)\, \mu_{exc}(n) 
+ \left(\!\sum_{j=N_Q+1}^\infty
 \frac{1}{j^5}\right)\frac{3e^2Q_{zz}}{a^5} \int
 d\mathbf{r}_\perp \, |W_m|^2(\rho) \left( 2z^2-\rho^2 \right).
\label{modkohneq2}
\ea
The total energy per unit cell [see Eq.~(\ref{energyeq})] is given by
\ba
E_\infty & = & \frac{a}{2\pi} \sum_{m\nu} \int_{I_{m\nu}} dk \,
\varepsilon_{m\nu}(k) - \frac{e^2}{2} \int \!\!
\int_{|z|<a/2,\,|z'|<a(N_Q+1/2)} d\mathbf{r} d\mathbf{r}' \,
\frac{n(\mathbf{r}) n(\mathbf{r}')}{|\mathbf{r} - \mathbf{r}'|}
\nonumber\\
& & + \int_{|z|<a/2} d\mathbf{r} \, n(\mathbf{r})
[\varepsilon_{\rm exc}(n)-\mu_{\rm exc}(n)] + \sum_{j=1}^{N_Q}
\frac{Z^2e^2}{ja} - \left(\sum_{j=N_Q+1}^\infty
\frac{1}{j^5}\right)\frac{3e^2}{2}\frac{Q_{zz}^2}{a^5} \,.  
\label{energyeq2}
\ea
In practice, we have found that accurate results are obtained for the 
energy of the chain even with $N_Q=1$ (i.e., only the primary cell and
its nearest neighbors are treated exactly and more distant cells are treated
using quadrupole approximation).

Details of our method used in computing the various integrals 
above and solving the Kohn-Sham equations self-consistently are given in the
Appendix.

%%%%%%%%%%%%%%%%%%%%%%%%%%%%%%%%%%%%%%%%%%%%%%%%%%%%%%%%%%%%%
\subsection{The electron band structure shape and occupations}
\label{subsec:band}

As discussed above, the electron orbitals in the chain are specified by 
three quantum numbers: $m,\nu,k$. While $m,\nu$ are discrete,
$k$ is continuous. In the ground state, the electrons will occupy the 
$(m\nu k)$ orbitals with the lowest energy eigenvalues 
$\varepsilon_{m\nu}(k)$. To determine the electron occupations and the 
total chain energy, it is necessary to calculate the $\varepsilon_{m\nu}(k)$ 
energy curves. Here we discuss the qualitative property of these energy curves
(i.e., the electron band structure) using the theory of 
one-dimensional periodic potentials (see, e.g., Ref.~\citep{ashcr76}).

Like the wave functions, the energy curves are periodic, with
$\varepsilon_{m\nu}(k+K) = \varepsilon_{m\nu}(k)$,
where $K$ is $2\pi/a$ multiplied by any integer.
The energy curves are also symmetric about the Bragg ``planes'' (``points''
in 1D) of the reciprocal lattice,
$\varepsilon_{m\nu}(K-k) = \varepsilon_{m\nu}(k)$.
Thus we can determine the entire band structure
of the electrons by calculating it between any two Bragg points. Since
we have chosen to limit our calculation to the first Brillouin zone
$k \in [-\pi/a,\pi/a]$, we only need to consider the domain 
$k \in [0,\pi/a]$.

\begin{figure}
\includegraphics[width=6.5in]{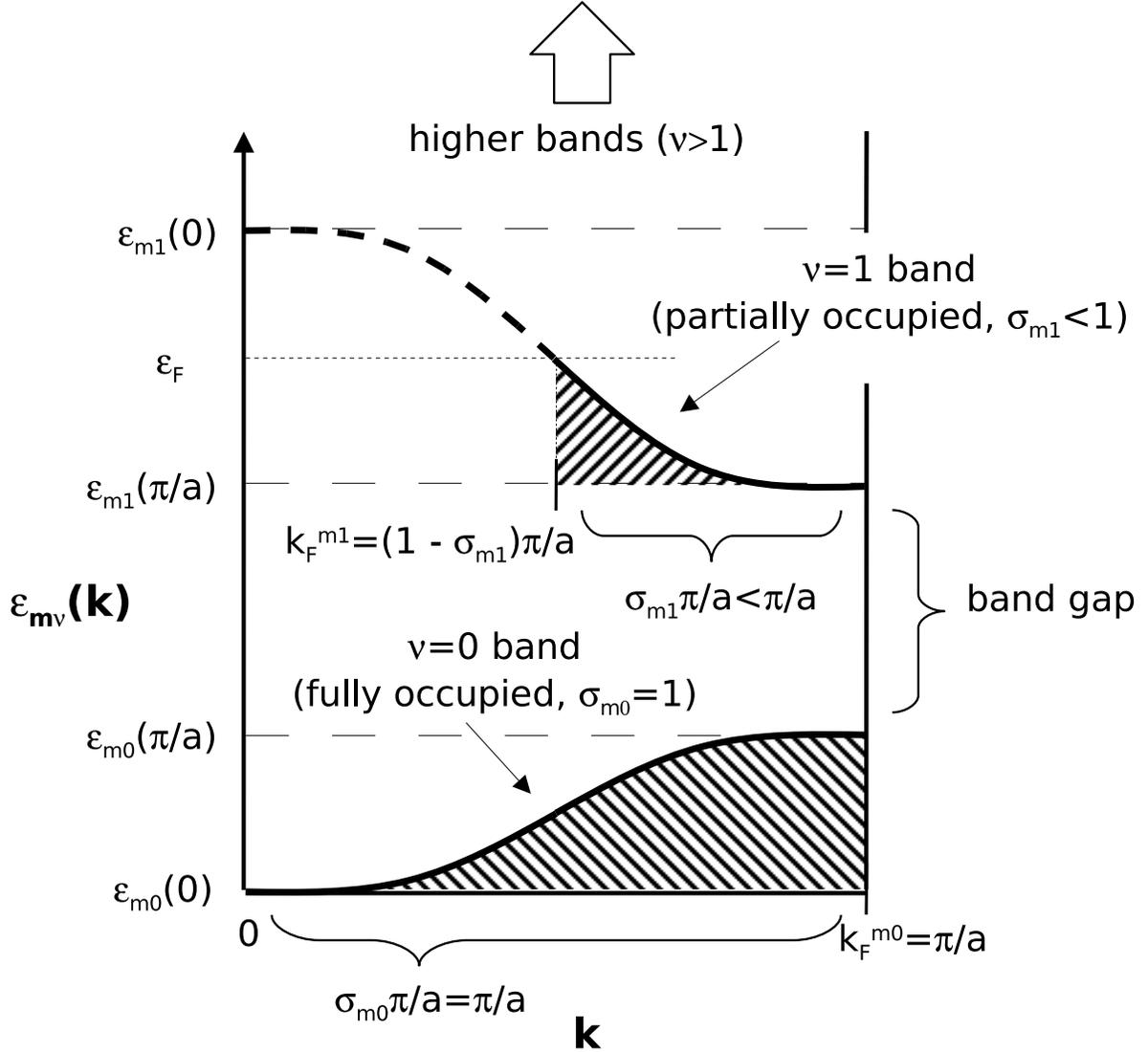}
\caption{A schematic diagram showing the electron band structure
for a particular $m$ value. In this example, the first band
($\nu=0$) is fully occupied ($\sigma_{m0}=1$) 
while the second band ($\nu=1$) is partially filled ($\sigma_{m1}<1$).}
\label{bandfig}
\end{figure}

For a given $m$, the energy curves lie in bands which do not overlap 
and increase in energy with increasing $\nu$ (see Fig.~\ref{bandfig}).
These bands are bounded by the energy values at the Bragg points, such
that in each band the energy increases/decreases monotonically between
the two points. The direction of this growth alternates with $\nu$:
For the $\nu=0$ band, the energy is at a minimum for $k=0$ and
increases to a local maximum at $k=\pi/a$; for the $\nu=1$ band, the
energy curve is at a minimum for $k=\pi/a$ and grows to a maximum at
$k=0$, etc. These properties are depicted in Fig.~\ref{bandfig}.

%%%%%%%%%%%%%%%%%%%%%
%%%%%%%%%%%%%%%%%%%%%

Also shown in the figure is the Fermi level energy $\varepsilon_F$ of the
electrons in the infinite chain. The electrons occupy all orbitals
$(m\nu k)$ with energy less than $\varepsilon_F$. For each $(m\nu)$ band, 
we define the occupation parameter $\sigma_{m\nu}$, which gives the number of
electrons that occupy this band per unit cell [i.e.,  the number of electrons
that occupy the $(m\nu)$ band in the whole chain is $\sigma_{m\nu}N$].
Since the maximum possible number of electrons in each $(m\nu)$ band is $N$, 
we have $\sigma_{m\nu} \le 1$. Because there are $ZN$ electrons total in 
the chain, these occupation numbers are subject to the constraint
\be
\sum_{m\nu} \sigma_{m\nu} = Z \,.
\label{constrainteq}
\ee
It is also useful to define for each $(m\nu)$ level the Fermi wave number
$k_F^{m\nu}$, such that the electrons fill up all allowed orbitals
between the minimum-energy Bragg point ($k=0$ for even $\nu$ and
$k=\pi/a$ for odd $\nu$) and $k_F^{m\nu}$. The occupied $k$'s
are therefore
\be
k \in \left[0,\sigma_{m\nu}\frac{\pi}{a}\right] \equiv 
\left[0,k_F^{m\nu}\right]
\ee
for even $\nu$, and
\be
k \in \left[(1-\sigma_{m\nu})\frac{\pi}{a},\frac{\pi}{a}\right] \equiv 
\left[k_F^{m\nu},\frac{\pi}{a}\right]
\ee
for odd $\nu$, plus the corresponding reflection about the Bragg point
$k=0$. For a completely filled band (as illustrated in
Fig.~\ref{bandfig} for the $\nu=0$ band), $\sigma_{m\nu}=1$ and
$k_F^{m\nu}=\pi/a$ (for $\nu=$ even) or $0$ (for $\nu=$ odd); for a
partially filled band (the $\nu=1$ band in Fig.~\ref{bandfig}),
\be
\varepsilon_{m\nu}(k_F^{m\nu}) = \varepsilon_F \,.
\label{mineq}
\ee
With the allowed $k$ values specified, the $k$ integration domain in Eqs.
(\ref{eq:density}), (\ref{eq:density2}), (\ref{energyeq}) and (\ref{energyeq2})
is given by
\be
\int_{I_{m\nu}}dk\Rightarrow
\left\{
\begin{array}{ll}
2\int_0^{k_F^{m\nu}}dk, & \mbox{ $\nu$ even,} \\
2\int_{k_F^{m\nu}}^{\pi/a}dk \,, & \mbox{ $\nu$ odd.}
\end{array}
\right.
\ee

Note that the Fermi level energy $\varepsilon_F$ and various
occupation numbers $\sigma_{m\nu}$ must be calculated
self-consistently. In principle, they should be determined by
minimizing the total energy with respect to $\sigma_{m\nu}$ subject to
the constraint Eq.~(\ref{constrainteq}), i.e.,
\be
\frac{\delta}{\delta \sigma_{m\nu}}\left[E[n;\sigma_{m\nu}] -
\varepsilon_F\left(\sum_{m\nu}\sigma_{m\nu}-Z\right)\right] = 0 \,.
\label{eq:min}\ee
Since
\be
\frac{\partial n(\mathbf{r})}{\partial \sigma_{m\nu}} = \pm\frac{\pi}{a}
\frac{\partial n(\mathbf{r})}{\partial k_F^{m\nu}} = |W_m|^2(\rho) 
|f_{m\nu k_F^{m\nu}}(z)|^2,
\ee
Eq.~(\ref{eq:min}) yields
\be 
\left[ -\frac{\hbar^2}{2m_e}\frac{d^2}{dz^2} 
+{\bar V}_{\rm eff}(z)\right]
f_{m\nu k_F^{m\nu}}(z) = \varepsilon_F f_{m\nu k_F^{m\nu}}(z) \,.
\ee
Comparing this to Eq.~(\ref{kohneq}), we find
$\varepsilon_{m\nu}(k_F^{m\nu}) = \varepsilon_F$, which is Eq.~(\ref{mineq}).
This shows that using Eq.~(\ref{mineq}) to find $\varepsilon_F$ 
minimizes the total energy of the system.

%%%%%%%%%%%%%%%%%%%%%
\begin{figure}
\includegraphics[width=6.5in]{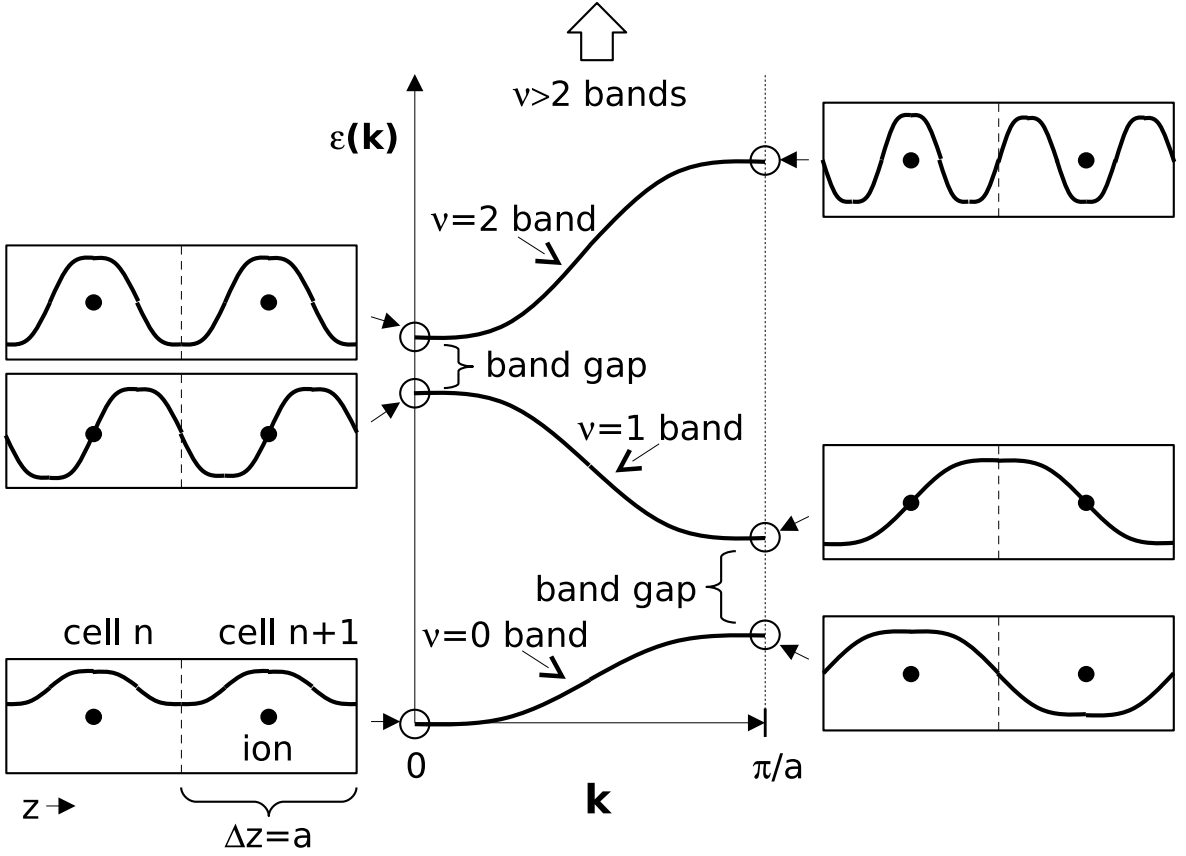}
\caption{A schematic diagram showing the shapes of the longitudinal
wave functions of electrons in different bands at $k=0$ and $k=\pi/a$.}
\label{wshapes}
\end{figure}

%%%%%%%%%%%%%%%%%%%%%%%%%%%%%%%%%
\subsection{The complex longitudinal wave functions}
\label{subsec:complex}

The longitudinal electron wave function $f_{m\nu k}(z)$ 
satisfies the Kohn-Sham equations (\ref{kohneq}) subject to the
periodicity condition Eq.~(\ref{periodeq}), or equivalently, the cell
boundary condition
\be
f_{m\nu k}(a/2) = e^{ika}f_{m\nu k}(-a/2) \,.
\label{bceq}
\ee
Since the electron density distribution $n(\mathbf{r})$
is periodic across each cell and symmetric about each ion, the following
boundary condition is also useful:
\be
|f_{m\nu k}(z)|'|_{z=a/2} = |f_{m\nu k}(z)|'|_{z=-a/2} = 0.
\label{bceq2}
\ee

Due to the complex boundary condition Eq.~(\ref{bceq}), the
wave function $f_{m\nu k}$ is complex for general $k$'s.  The
exceptions are $k=0$ and $k=\pi/a$: For $k=0$, the boundary condition
becomes $f_{m\nu k}(a/2) = f_{m\nu k}(-a/2)$, and for $k=\pi/a$ we
have $f_{m\nu k}(a/2) = -f_{m\nu k}(-a/2)$. Thus for $k=0$ and
$\pi/a$, we can choose the longitudinal wave functions to be real. The
general shapes of these wave functions (for different bands) are
sketched out in Fig.~\ref{wshapes}.  We see that at the Bragg points,
between the two states with the same number of nodes, the one that is
more concentrated near the ion has lower energy than the other state;
this difference gives rise to the band gap. The $k=0,\pi/a$
eigenvalues $\varepsilon_{m\nu}$ and eigenfunctions can be calculated
in the domain $0<z<a/2$ with the boundary condition $f_{m\nu k}(0)=0$
or $f'_{m\nu k}(0)=0$.

The electron wave functions for general $k$'s are more difficult to
compute as they have complex boundary conditions. Our procedure for
calculating these wave functions and their corresponding electron
energies is as follows: For each energy band $(m\nu)$, the electron
eigenstates at $k=0$ and $k=\pi/a$ are first found (see above). For
every energy between $\varepsilon_{m\nu}(k=0)$ and
$\varepsilon_{m\nu}(k=\pi/a)$, we find the wave function that solves
the Kohn-Sham equation while satisfying the symmetric/periodic density
condition Eq.~(\ref{bceq2}). More precisely, we choose $f=1$ (up to a
normalization constant) and guess $f'=i\,g$ (where $g$ is a real
number) at $z=a/2$ (thus $|f|'=0$ is satisfied at $z=a/2$), and then
integrate the Kohn-Sham equation to $z=-a/2$; we adjust $g$ so that
$|f|'=0$ is satisfied at $z=-a/2$.
Once the wave function is obtained, we determine its $k$ value from the
Bloch boundary condition Eq.~(\ref{bceq}). Through this method we find
$\varepsilon_{m\nu}(k)$ as a function of $k$ for each $(m\nu)$ band.

\begin{figure}
\includegraphics[width=6.5in]{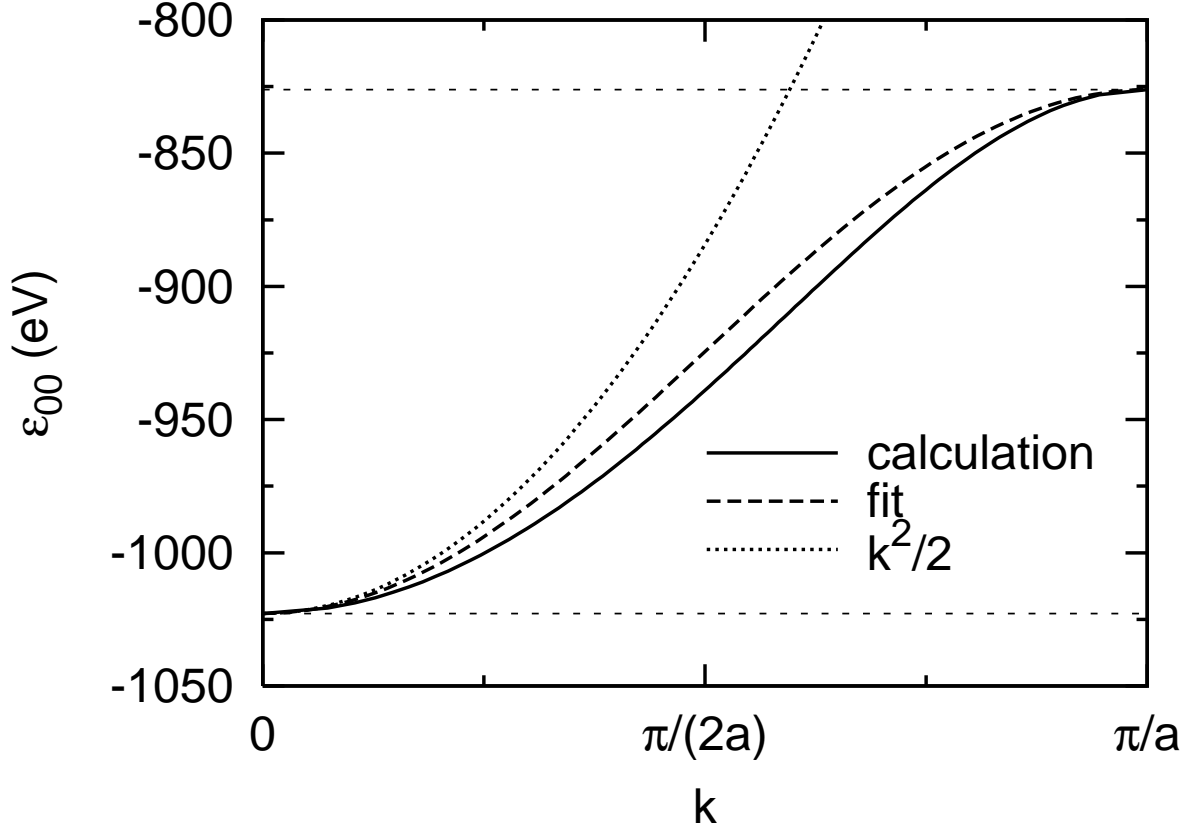}
\caption{The electron energy of the $(m,\nu)=(0,0)$ band for the
carbon infinite chain at $B_{12}=1$. The tight-binding model fit for
this level is shown as a dashed line [see Eq.~(\ref{tightbeq})], and the dotted line shows the free electron result $\varepsilon_{00}(k)-\varepsilon_{00}(0)=k^2/2$.}
\label{Cband}
\end{figure}

\begin{figure}
\includegraphics[width=6.5in]{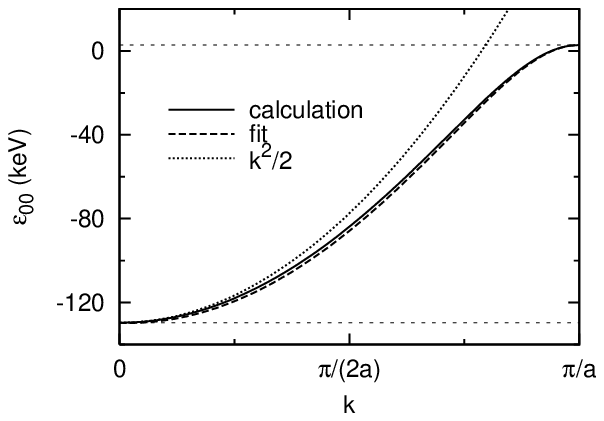}
\caption{The electron energy of the $(m,\nu)=(0,0)$ band for the iron
infinite chain at $B_{12}=2000$. The weak-periodic-potential model fit
for this level is shown as a dashed line [see Eq.~(\ref{weakbeq})],
and the dotted line shows the free electron result
$\varepsilon_{00}(k)-\varepsilon_{00}(0)=k^2/2$.}
\label{Feband}
\end{figure}

Some examples of our computed $\varepsilon_{m\nu}(k)$ are shown in
Figs.~\ref{Cband} and \ref{Feband}. To show that our calculations are
consistent with theoretical models, we have included several model
fits for the electron energy curves: the tight-binding fit in
Fig.~\ref{Cband}, which has the form
\be
\varepsilon_{m\nu}(k) \simeq c_1 + c_2 \cos(ka)
\label{tightbeq}
\ee
[see Ref.~\citep{ashcr76}, Eq.~(10.19)], and the
weak-periodic-potential fit in Fig.~\ref{Feband}, which has the form
\be
\varepsilon_{m\nu}(k) \simeq c_1 + \frac{1}{2}[k^2/2+(2\pi/a-k)^2/2]
- \frac{1}{2}\{[(2\pi/a-k)^2/2-k^2/2]^2+c_2^2\}^{1/2}
\label{weakbeq}
\ee
[see Ref.~\citep{ashcr76}, Eq.~(9.26)]. The constants $c_1$ and $c_2$
in the formulas are fit to the two endpoints of the energy curves,
$\varepsilon_{m\nu}(0)$ and $\varepsilon_{m\nu}(\pi/a)$.  The
tight-binding model fits well for the most-tightly-bound electron
bands in our calculations, while the weak-periodic-potential model
fits well for all of the other bands.  Note that for $k \ll \pi/a$,
the electron energy can be approximately fit by
$\varepsilon_{m\nu}(k)=\varepsilon_{m\nu}(0)+k^2/2$,
as would be the case if the wave functions were of the form
$f_{m\nu}(z)e^{ikz}$ --- this is the ansatz adopted by 
\citet{neuhauser87} in their Hartree-Fock calculations.
But obviously for larger $k$,
this is a rather bad approximation. We suggest that approximate
treatment in the band structure may account for a large part of the
discrepancies among cohesive energy results in previous works.
For example, the disagreement between
Ref.~\citep{jones85} 
[where $\varepsilon_{m\nu}(k)$ was calculated for a few values of 
$k$ and then fit to a simple expression]
and Ref.~\citep{neuhauser87} (where a $k^2$ dependence for 
the electron energy was assumed) on whether or not carbon is bound at 
$B_{12}=5$ is due to the band structure model, not to the fact 
the former used the density functional theory while the latter
used the Hartree-Fock method.

%%%%%%%%%%%%%%%%%%%%

%%%%%%%%%%%%%%%%%%%%

%%%%%%%%%%%%%%%%%%%%%%%%%%%%%%%%%%%%%%%%%%%%%%%%%%%%%%%%%%%%%%%%%%
\section{Results: One-dimensional chains}
\label{sec:result-1d}

In this section we present our results for hydrogen, helium, carbon,
and iron infinite chains at various magnetic field strengths between
$B=10^{12}$~G and $2\times10^{15}$~G\@. For each chain, data is given in
tabular form for the ground-state energy (per unit cell) $E_\infty$,
the equilibrium ion separation $a$, and the electron Fermi level
energy $\varepsilon_F$ (the electron work function is
$W=|\varepsilon_F|$). We provide relevant information for the
electron occupations in different bands, such as the number of Landau
orbitals and the number of fully occupied bands (see below for
specific elements). We also give the ground-state energy of the
corresponding atom, $E_a$, so that the cohesive energy of each chain
can be obtained, $Q_\infty=E_a-E_\infty$.

For each chain and atom we provide numerical scaling
relations for the ground-state energy and Fermi level energy as a function
of the magnetic field, in the form of scaling exponents $\beta$ and
$\gamma$, with 
\be
E_a,~E_\infty \propto B^\beta,\qquad \varepsilon_F \propto B^\gamma.
\ee
We also give the rescaled, dimensionless energy $\bar E_\infty$,
and equilibrium ion separation $\bar a$ defined by
[see Eq.~(\ref{chainscale})]
\be
E_\infty \simeq \bar{E}_\infty\, Z^{9/5} b^{2/5}~{\rm a.u.},\qquad
a \simeq \bar{a}\, Z^{1/5} b^{-2/5}~{\rm a.u.}.
\label{eq:scale}
\ee

We shall see that the scaling relations in Eq.~(\ref{eq:scale}) with
$\bar{E}_\infty \simeq$ const.\ and $\bar{a} \simeq$ const.\ represent
a reasonable approximation to our numerical results, although such
scaling formulae are not accurate enough for calculating the cohesive
energy $Q_\infty = E_a - E_\infty$. However, Eq.~(\ref{fermiscale}) or
Eq.~(\ref{fermiscale3D}) for the Fermi level energy based on the
uniform gas model is not a good representation of our numerical
results. In paper I \citep{medin06a} we have shown that as $N$
increases, the energy per atom in the H$_N$ (or He$_N$, C$_N$, Fe$_N$)
molecule, $E_N/N$, gradually approaches a constant value. The infinite
chain ground-state energy $E_\infty$ found in the present paper is
consistent with the large-$N$ molecule ground-state energy limit
$E_N/N$ obtained in paper I (see the related figures in the following
subsections). Since finite molecules and infinite chains involve
completely different treatments of the electron states, the
consistency of $E_\infty$ and $E_N/N$ provides an important check of
the validity of our calculations.

Other comparisons can be made between the infinite chains and finite
molecules. For example, our results of ion separation $a$ and scaling
constant $\beta$ are consistent between infinite chains and finite
molecules.  Also, we find that if the isolated atom has electrons in
$\nu=0$ and $\nu=1$ orbitals, then the corresponding infinite chain
will have electrons in $\nu=0$ and $\nu=1$ bands; if the isolated atom
only has electrons in $\nu=0$ orbitals, the corresponding infinite
chain will have electrons only in $\nu=0$ bands.

We have compared our cohesive energy results with those of other
work, whenever available.
These comparisons are presented in the following subsections.

%%%%%%%%%%%%%%%%%%%%%%%%%%%%%%%%%
\subsection{Hydrogen}

\begin{figure}
\includegraphics[width=6.5in]{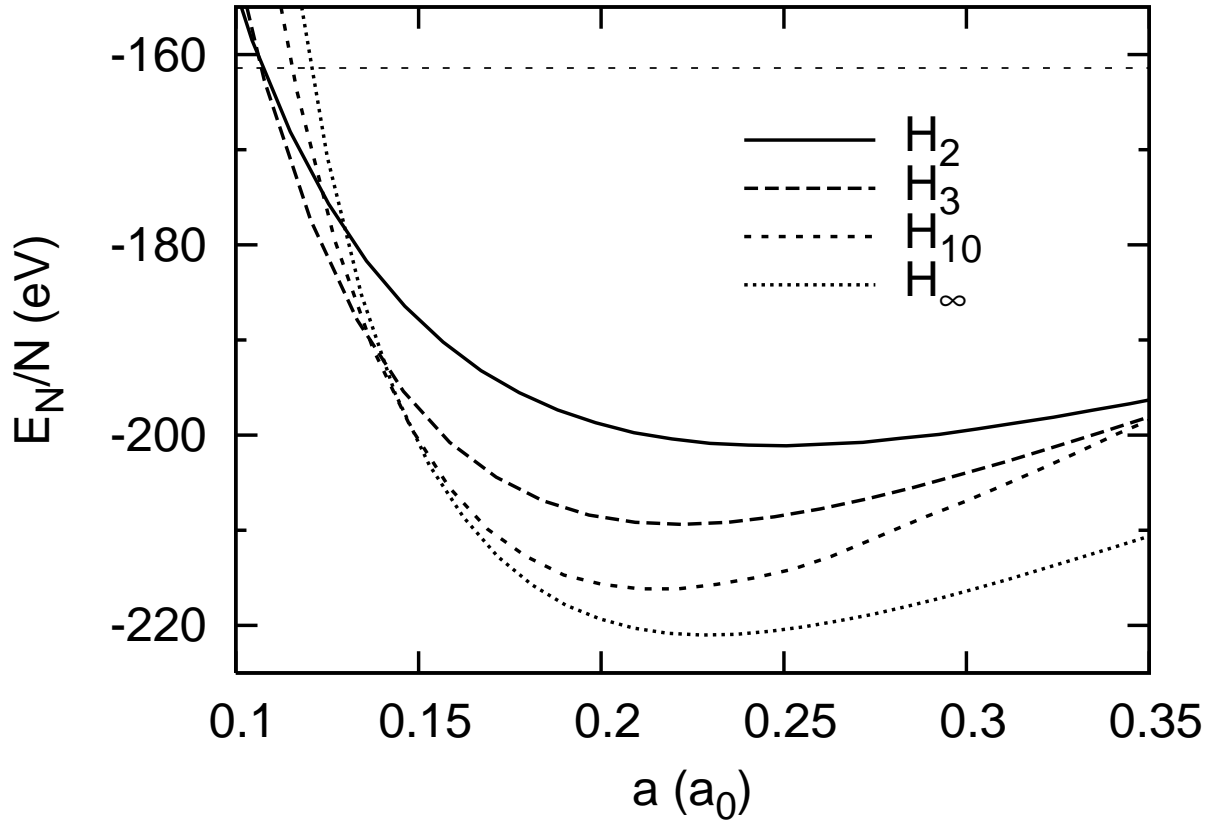}
\caption{The energies (per atom or cell) of various H molecules and 
infinite chain as a function of ion separation $a$ at $B_{12}=1$. 
The results of finite molecules are based on paper I \citep{medin06a}.
The energy of the H atom is shown as a horizontal line at $-161.4$~eV.}
\label{HChnfig}
\end{figure}

\begin{figure}
\includegraphics[width=6.5in]{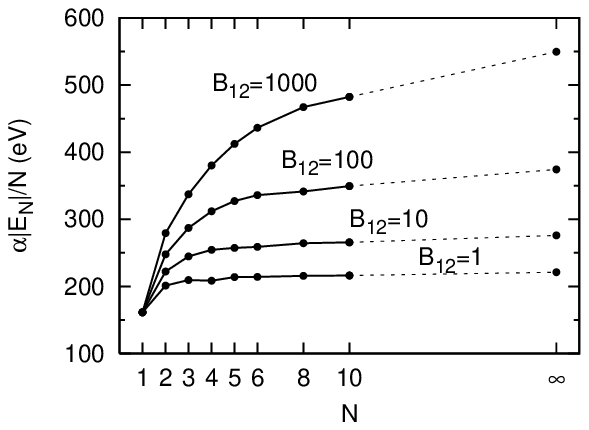}
\caption{The molecular energy per atom, $|E_N|/N$, for the H$_N$
molecule, as a function of $N$ at several different field strengths.
The results of finite molecules are based on paper I \citep{medin06a}.
As $N$ increases, $E_N/N$ asymptotes to $E_\infty$.  To facilitate
plotting, the values of $|E_1|$ (atom) at different magnetic field
strengths are normalized to the value at $B_{12}=1$, $161.4$~eV\@. This
means that $\alpha = 1$ for $B_{12}=1$, $\alpha = 161.4/309.5$ for
$B_{12}=10$, $\alpha = 161.4/540.3$ for $B_{12}=100$, and $\alpha =
161.4/869.6$ for $B_{12}=1000$.}
\label{HMolfig}
\end{figure}

\begin{figure}
\includegraphics[width=6.5in]{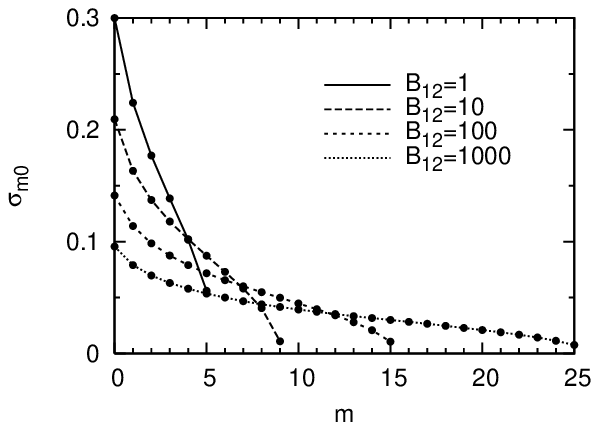}
\caption{The occupation numbers of each $m$ level of hydrogen infinite
chains, for various magnetic field strengths. The data points are
plotted over the curves to show the discrete nature of the $m$
levels. Note that only the $\nu=0$ bands are occupied by the electrons.}
\label{Hsigma}
\end{figure}

Our numerical results for H are given in Table~\ref{Htable}. Examples
of the energy curves of various H$_N$ molecules and H$_\infty$ at
$B_{12}=1$ are depicted in Fig.~\ref{HChnfig}. The minimum of each
energy curve determines the equilibrium ion separation in the
molecule/chain.  Figure~\ref{HMolfig} compares the molecular and
infinite chain energies at various field strengths, and shows that as
$N$ increases, the energy per atom in the H$_N$ molecule asymptotes to
$E_\infty$. Figure~\ref{Hsigma} gives the occupation number
$\sigma_{m0}$ of different Landau orbitals at various field
strengths. Only the $\nu=0$ bands are occupied, none of these are
completely filled ($\sigma_{m0}<1$), and the $\nu\ge 1$ bands are
empty ($\sigma_{m1}=0$). We see that as $B$ increases, the electrons
spread into more Landau orbitals, thus the number of $m$ states
occupied by the electrons ($n_m$ in Table~\ref{Htable})
increases. Approximately, since the chain radius $R\propto b^{-2/5}$
and $R\sim (2n_m-1)^{1/2}/b^{1/2}$ (the electrons occupy the
Landau orbitals with $m=0,1,2,\ldots, n_m-1$), we have $n_m\propto
b^{1/5}$.  Table~\ref{Htable} shows that for $B_{12}\agt 10$ our
results for $E_\infty$ and $a$ are well fit by
\be
E_\infty \simeq -529\,B_{13}^{0.374}~{\rm eV},\qquad
a = 0.091\,B_{13}^{-0.40}~{\rm }a_0
\ee
[where $B_{13}=B/(10^{13}~{\rm G})$], similar to the scaling of
Eq.~(\ref{eq:scale}).  The electron work function $W=|\varepsilon_F|$
does not scale as Eq.~(\ref{fermiscale}), but is a fraction of the
ionization energy of the H atom, $|E_a|$. Note that $|E_a|$ is not
well fit by a power law ($\propto B^\beta$), but is well described by
$|E_a|\propto (\ln b)^2$ (accurate fitting formulae for $|E_a|$ are
given in, e.g., Ref.~\citep{ho03}).

At $B_{12}=1,10,100$, we find cohesive energies of
$Q_\infty=E_a-E_\infty=59.6$,~$219.7$,~$712.7$~eV (see
Table~\ref{Htable}). At those same fields, \citet{lai92} find cohesive
energies of $28.9$,~$141$,~$520$~eV\@. At $B_{12}=0.94$,
\citet{relovsky96} find a cohesive energy of $47.1$~eV\@. We expect
our H calculation (and that of Ref.~\citep{relovsky96}) to
overestimate the cohesive energy since an exchange-correlation
functional is used in the chain calculation while none is required for
the H atom. But we also expect the result obtained in Ref.~\citep{lai92} to
somewhat underestimate the cohesive energy since a uniform
(longitudinal) electron density was assumed.

\begin{table}
\caption{The ground-state energy (per unit cell) $E_\infty$ (in units of
eV), ion separation $a$ (in units of Bohr radius $a_0$), the number of
occupied Landau levels $n_m$, 
and the Fermi level energy $\varepsilon_F$ (in eV) of 
1D infinite chains of hydrogen, over a range of magnetic field strengths. 
The ground-state energy of individual hydrogen atoms, $E_a$ (in units of eV), 
is also provided for reference.
The dimensionless energy $\bar E_\infty$ and ion separation $\bar a$
are calculated using Eq.~(\ref{eq:scale}).
The scaling exponents $\beta$ and $\gamma$, defined by 
$E_a,~E_\infty \propto B^\beta$, and $\varepsilon_F \propto B^\gamma$,
are calculated over the three magnetic field ranges provided in the table:
$B_{12}=1-10$, $10-100$, $100-1000$ (the exponent in the
$B_{12}=1$ row corresponds to the fit over $B_{12}=1-10$, etc.).
The occupation of different $(m\nu)$ bands is designated by the number
$n_m$: the electrons occupy Landau orbitals with $m=0,1,2,\ldots,n_m-1$,
all in the $\nu=0$ band; see Fig.~\ref{Hsigma}.}
\label{Htable}
\centering
\begin{tabular}{c | r@{}l c | r@{}l l c l c c c c}
\hline\hline
 & \multicolumn{3}{| c |}{H} & \multicolumn{9}{| c}{H$_\infty$} \\
$B_{12}$ & \multicolumn{2}{| c}{$\quad E_a\quad $} & $\beta$ & 
\multicolumn{2}{| c}{
$\quad E_\infty\quad $} & \multicolumn{1}{c}{$\bar{E}_\infty$} 
& $\beta$ & \multicolumn{1}{c}{$a$} & \multicolumn{1}{c}{$\bar{a}$} & $n_m$ & $\varepsilon_F$ & $\gamma$ \\
\hline
1 & -161&.4 & 0.283 & -221&.0 & -0.721 & 0.379 & 0.23 & 2.6 & 6 & -85.0 & 0.28 \\
\hline
10 & -309&.5 & 0.242 & -529&.2 & -0.688 & 0.374 & 0.091 & 2.6 & 10 & -165 & 0.27 \\
\hline
100 & -540&.3 & 0.207 & -1253&.0 & -0.648 & 0.374 & 0.037 & 2.6 & 16 & -311 & 0.26 \\
\hline
1000 & -869&.6 & - & -2962& & -0.610 & - & 0.0145 & 2.6 & 26 & -571 & - \\
\hline\hline
\end{tabular}
\end{table}

%%%%%%%%%%%%%%%%%%%%%%%%%%%%%%%%%%%%%%%%%%%%%%%%%%%%%%%%%%%%%%%%%%
\subsection{Helium}

\begin{figure}
\includegraphics[width=6.5in]{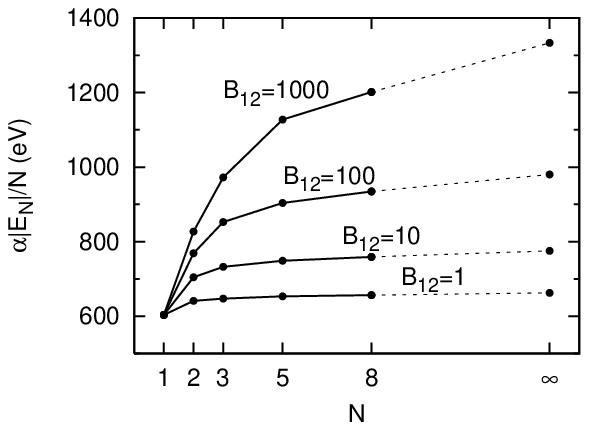}
\caption{The molecular energy per atom, $|E_N|/N$, for the He$_N$
molecule, as a function of $N$ at several different field strengths.
The results of finite molecules are based on paper I \citep{medin06a}.
As $N$ increases, $E_N/N$ asymptotes to $E_\infty$.  To facilitate
plotting, the values of $|E_1|$ (atom) at different magnetic field
strengths are normalized to the value at $B_{12}=1$, $603.5$~eV\@. This
means that $\alpha = 1$ for $B_{12}=1$, $\alpha = 603.5/1252.0$ for
$B_{12}=10$, $\alpha = 603.5/2385$ for $B_{12}=100$, and $\alpha =
603.5/4222$ for $B_{12}=1000$.}
\label{HeMolfig}
\end{figure}

\begin{figure}
\includegraphics[width=6.5in]{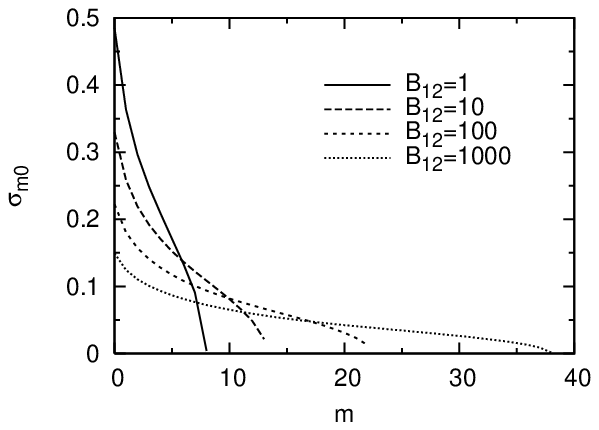}
\caption{The occupation numbers of each $m$ level of infinite He
chains, for various magnetic field strengths.  Only the $\nu=0$ bands
are occupied by the electrons.  Note that for $B_{12}=1$, the $m=8$
orbital has a rather small occupation, $\sigma_{80}\simeq 0.006$; if
$\varepsilon_F$ were slightly more negative, this orbital would be
completely unoccupied.}
\label{Hesigma}
\end{figure}

Our numerical results for He are given in Table~\ref{Hetable}.
Figure~\ref{HeMolfig} compares the molecular and infinite chain energies
at various field strengths, and shows that as $N$ increase, the energy
per atom in the He$_N$ molecule approaches $E_\infty$ for the infinite
chain. Figure~\ref{Hesigma} gives occupation number $\sigma_{m0}$ of
different Landau orbitals at various field strengths. As in the case
of H, only the $\nu=0$ bands are occupied, and the number of Landau
states required ($n_m$ in Table~\ref{Hetable}) increases with
increasing $B$, with $n_m\propto Z^{2/5}b^{1/5}$. Table~\ref{Hetable}
shows that for $B_{12}\agt 10$,
\be
E_\infty \simeq -1252\,B_{13}^{0.382}~{\rm eV},\qquad
a = 0.109\,B_{13}^{-0.40}~{\rm }a_0\,,
\ee
similar to the scaling of Eq.~(\ref{eq:scale}). The electron work
function $W=|\varepsilon_F|$ does not scale as Eq.~(\ref{fermiscale}),
but is a fraction of the ionization energy: Using a Hartree-Fock code
(e.g., Ref.~\citep{lai92}) we find that at $B_{12}=1,10,100,1000$ the
He atomic energies are $-575.5$,~$-1178.0$,~$-2193$,~$-3742$~eV\@. The
He$^+$ (i.e., once-ionized He) energies at these field strengths are
$-416.2$,~$-846.5$,~$-1562.0$,~$-2638$~eV\@. Therefore, the ionization
energies of He at these field strengths are
$159.3$,~$331.5$,~$631$,~and~$1104$~eV, respectively.

At $B_{12}=1$, we find a cohesive energy of $58.9$~eV 
(see Table~\ref{Hetable}). At the same field, \citet{neuhauser87} 
(based on the Hartree-Fock model) find a cohesive energy of $25$~eV, 
and \citet{muller84} (based on variational methods) 
gives a cohesive energy of $50$~eV\@. At $B_{12}=0.94$, \citet{relovsky96} 
(based on density functional theory) find a cohesive energy of $56.6$~eV\@. 
At $B_{12}=5$ \citet{jones85} finds a cohesive energy of
$220$~eV, which is close to our value.
That our results agree best with those of
Refs.~\citep{relovsky96,jones85} is expected, as we used a similar
method to find the ground-state atomic and chain energies. Similar to
the finite He molecules (paper I), we expect our
density-functional-theory calculation to overestimate the cohesive
energy, but we also expect the result of Ref.~\citep{neuhauser87} to
underestimate $Q_\infty$.

\begin{table}
\caption{The ground-state energy (per unit cell) $E_\infty$ (in units
of eV), ion separation $a$ (in units of Bohr radius $a_0$), the number
of occupied Landau levels $n_m$, and Fermi level energy
$\varepsilon_F$ (in eV) of 1D infinite chains of helium, over a range
of magnetic field strengths.  The ground-state energy of individual He
atoms, $E_a$ (in units of eV), is also provided for reference (this is
based on the density-functional-theory calculation of
Ref.~\citep{medin06a}).  The dimensionless energy $\bar E_\infty$ and
ion separation $\bar a$ are calculated using Eq.~(\ref{eq:scale}).
The scaling exponents $\beta$ and $\gamma$, defined by $E_a,~E_\infty
\propto B^\beta$, and $\varepsilon_F \propto B^\gamma$, are calculated
over the three magnetic field ranges provided in the table:
$B_{12}=1-10$, $10-100$, $100-1000$ (the exponent in the $B_{12}=1$
row corresponds to the fit over $B_{12}=1-10$, etc.).  The occupation
of different $(m\nu)$ bands is designated by the number $n_m$: the
electrons occupy Landau orbitals with $m=0,1,2,\ldots,n_m-1$, all in
the $\nu=0$ band; see Fig.~\ref{Hesigma}.  Note that all of the He
atoms here also have electrons only in the $\nu=0$ states.}
\label{Hetable}
\centering
\begin{tabular}{c | r@{}l c | r@{}l l c l c c c c}
\hline\hline
 & \multicolumn{3}{| c |}{He} & \multicolumn{9}{| c}{He$_\infty$} \\
$B_{12}$ & \multicolumn{2}{| c}{$E_a$} & $\beta$ 
& \multicolumn{2}{| c}{$E_\infty$} & \multicolumn{1}{c}{$\bar{E}_\infty$} 
& $\beta$ & \multicolumn{1}{c}{$a$} & \multicolumn{1}{c}{$\bar{a}$} & $n_m$ & $\varepsilon_F$ & $\gamma$ \\
\hline
1 & -603&.5 & 0.317 & -662&.4 & -0.621 & 0.385 & 0.28 & 2.7 & 9 & -85.0 & 0.29 \\
\hline
10 & -1252&.0 & 0.280 & -1608&.0 & -0.600 & 0.382 & 0.109 & 2.7 & 14 & -167 & 0.27 \\
\hline
100 & -2385& & 0.248 & -3874& & -0.575 & 0.382 & 0.043 & 2.7 & 23 & -310 & 0.26 \\
\hline
1000 & -4222& & - & -9329& & -0.552 & - & 0.0175 & 2.7 & 39 & -568 & - \\
\hline\hline
\end{tabular}
\end{table}

%%%%%%%%%%%%%%%%%%%%%%%%%%%%%%%%%%%%%%%5
\subsection{Carbon}

\begin{figure}
\includegraphics[width=6.5in]{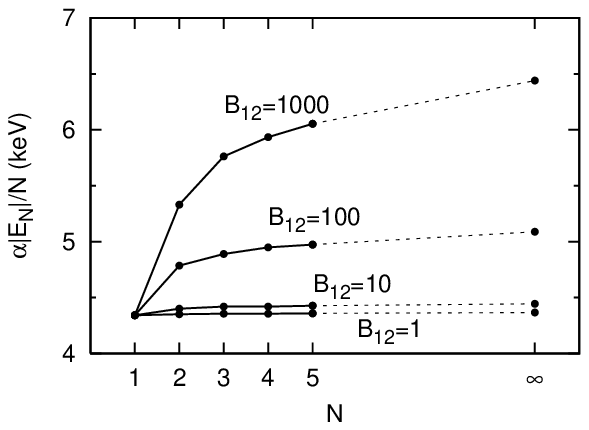}
\caption{The molecular energy per atom, $|E_N|/N$, for the C$_N$
molecule, as a function of $N$ at several different field strengths.
The results of finite molecules are based on paper I \citep{medin06a}.
As $N$ increases, $E_N/N$ asymptotes to $E_\infty$.  To facilitate
plotting, the values of $|E_1|$ (atom) at different magnetic field
strengths are normalized to the value at $B_{12}=1$, $4341$~eV\@.  This
means that $\alpha = 1$ for $B_{12}=1$, $\alpha = 4341/10\,075$ for
$B_{12}=10$, $\alpha = 4341/21\,360$ for $B_{12}=100$, and $\alpha =
4341/41\,330$ for $B_{12}=1000$.}
\label{CMolfig}
\end{figure}

\begin{figure}
\includegraphics[width=6.5in]{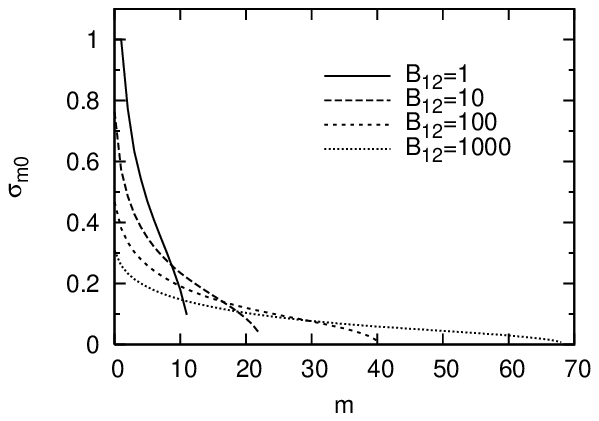}
\caption{The occupation numbers of each $m$ level of infinite
C chains, for various magnetic field strengths. 
Only the $\nu=0$ bands are occupied by the electrons.
Note that for $B_{12}=1$, the $m=0$ and $m=1$ bands are completely filled.}
\label{Csigma}
\end{figure}

Our numerical results for C are given in Table~\ref{Ctable}.
Figure~\ref{CMolfig} compares molecular and infinite chain
energies at various field strengths, showing that as $N$ increase, the
energy per atom in the C$_N$ molecule approaches $E_\infty$ for the
infinite chain. Figure~\ref{Csigma} gives the occupation number
$\sigma_{m0}$ of different Landau orbitals at various field
strengths. As in the case of H and He, only the $\nu=0$ bands are
occupied, although for C at $B_{12}=1$, the $m=0$ and $m=1$ bands
(both with $\nu=0$) are fully occupied (thus $n_f=2$ in
Table~\ref{Ctable}). The number of Landau states required ($n_m$ in
Table~\ref{Ctable}) increases with increasing $B$, approximately with
$n_m\propto Z^{2/5}b^{1/5}$. Table~\ref{Ctable} shows that for
$B_{12}\agt 10$,
\be
E_\infty \simeq -10\,300\,B_{13}^{0.387}~{\rm eV},\qquad
a = 0.154\,B_{13}^{-0.43}~{\rm }a_0\,.
\ee
Note that these expressions are more approximate than for H and He.
The electron work function $W=|\varepsilon_F|$ does not scale as
Eq.~(\ref{fermiscale}), but is a fraction of the ionization energy:
from paper I \citep{medin06a}, the ionization energies of C at
$B_{12}=1,10,100,1000$ are $174$,~$430$,~$990$,~and~$2120$~eV,
respectively.

At $B_{12}=10$, we find a cohesive energy of $240$~eV (see
Table~\ref{Ctable}).  At $B_{12}=8.5$, \citet{relovsky96} give a
cohesive energy of $240$~eV\@. At $B_{12}=5$ \citet{jones85} finds a
cohesive energy of $100$~eV; at the same field (using our scaling
relations), we find a cohesive energy of $100$~eV ($\pm
30$~eV). \citet{neuhauser87}, on the other hand, find that carbon is
not bound at $B_{12}=1$ or $5$.  This is probably due to the
approximate band structure ansatz adopted in Ref.~\citep{neuhauser87}
(see Sec.~\ref{subsec:complex}): for fully occupied bands, the
approximation that $\varepsilon_{m\nu}(k)$ increases as $k^2/2$ is
invalid and can lead to large error in the total energy of the chain.

%%%%%%%%%%%%%%%%%
\begin{table}
\caption{The ground-state energy (per unit cell) $E_\infty$ (in units
of eV), ion separation $a$ (in units of Bohr radius $a_0$), electron
occupation numbers ($n_m,n_f$), and Fermi level energy $\varepsilon_F$
(in eV) of 1D infinite chains of carbon, over a range of magnetic
field strengths. The ground-state energy of individual C atoms, $E_a$
(in units of eV), is also provided for reference (this is based on the
density-functional-theory calculation of Ref.~\citep{medin06a}). The
dimensionless energy $\bar E_\infty$ and ion separation $\bar a$ are
calculated using Eq.~(\ref{eq:scale}). The scaling exponents $\beta$
and $\gamma$, defined by $E_a,~E_\infty \propto B^\beta$, and
$\varepsilon_F \propto B^\gamma$, are calculated over the three
magnetic field ranges provided in the table: $B_{12}=1-10$, $10-100$,
$100-1000$ (the exponent in the $B_{12}=1$ row corresponds to the fit
over $B_{12}=1-10$, etc.). The occupation of different $(m\nu)$ bands
is designated by the notation $(n_m,n_f)$: the electrons occupy
Landau orbitals with $m=0,1,2,\ldots,n_m-1$, all with $\nu=0$; the
number of fully occupied ($\sigma_{m\nu}=1$) bands is denoted by
$n_f$; see Fig.~\ref{Csigma}. Note that all of the C atoms here also
have electrons only in the $\nu=0$ states.}
\label{Ctable}
\centering
\begin{tabular}{c | r c | r l c l c c c c}
\hline\hline
 & \multicolumn{2}{| c |}{C} & \multicolumn{8}{| c}{C$_\infty$} \\
$B_{12}$ & \multicolumn{1}{| c}{$E_a$} & $\beta$ & \multicolumn{1}{| c}
{$E_\infty$} & \multicolumn{1}{c}{$\bar{E}_\infty$} & $\beta$ & \multicolumn{1}{c}{$a$} & \multicolumn{1}{c}{$\bar{a}$} & $(n_m,n_f)$ & $\varepsilon_F$ & $\gamma$ \\
\hline
1 & -4341 & 0.366 & -4367 & -0.567 & 0.373 & 0.49 & 3.9 & (12,2) & -92.8 & 0.27 \\
\hline
10 & -10075 & 0.326 & -10315 & -0.533 & 0.385 & 0.154 & 3.1 & (23,0) & -173 & 0.25 \\
\hline
100 & -21360 & 0.287 & -25040 & -0.515 & 0.389 & 0.056 & 2.8 & (41,0) & -306 & 0.25 \\
\hline
1000 & -41330 & - & -61320 & -0.502 & - & 0.022 & 2.7 & (69,0) & -539 & - \\
\hline\hline
\end{tabular}
\end{table}

%%%%%%%%%%%%%%%%%%%%%%%%%%%%%%%%%%%%%%%%%%%%%%%%%%%%%
\subsection{Iron}

\begin{figure}
\begin{center}
\begin{tabular}{cc}
\resizebox{3in}{!}{\includegraphics{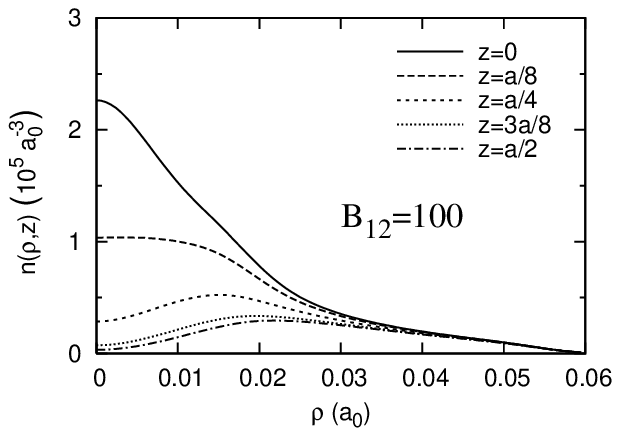}} &
\resizebox{3in}{!}{\includegraphics{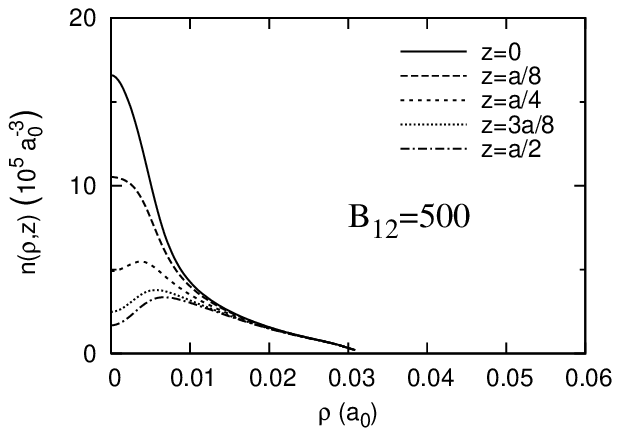}} \\
\resizebox{3in}{!}{\includegraphics{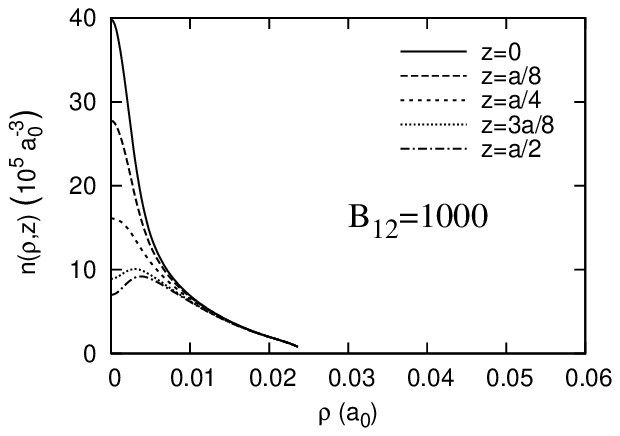}} &
\resizebox{3in}{!}{\includegraphics{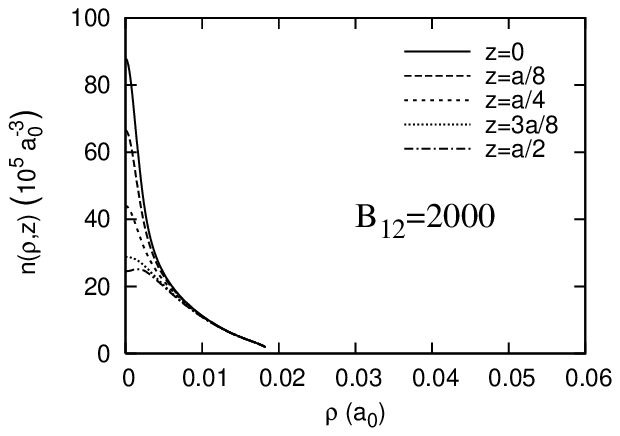}} \\
\end{tabular}
\caption{The density distribution of electrons in the
iron infinite chain at four different magnetic field
strengths (labeled on the graphs). The density is shown as a
function of $\rho$ for five equally spaced $z$ points from the center
of a cell ($z=0$) to the edge of that cell ($z=a/2$).}
\label{DensityCompfig}
\end{center}
\end{figure}

\begin{figure}
\begin{center}
\begin{tabular}{cc}
\resizebox{3in}{!}{\includegraphics{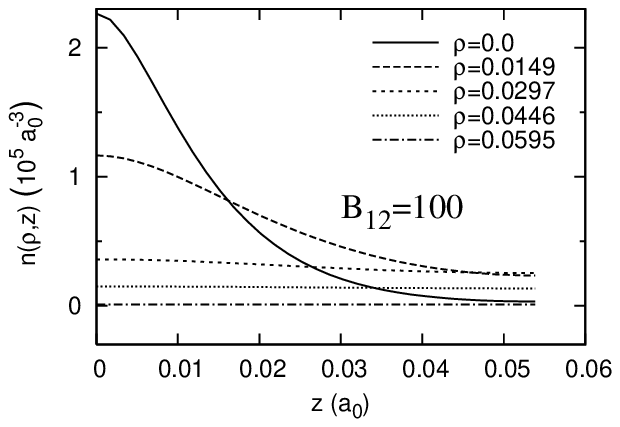}} &
\resizebox{3in}{!}{\includegraphics{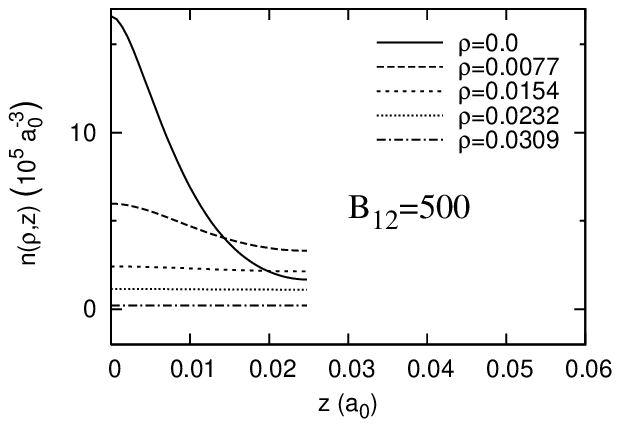}} \\
\resizebox{3in}{!}{\includegraphics{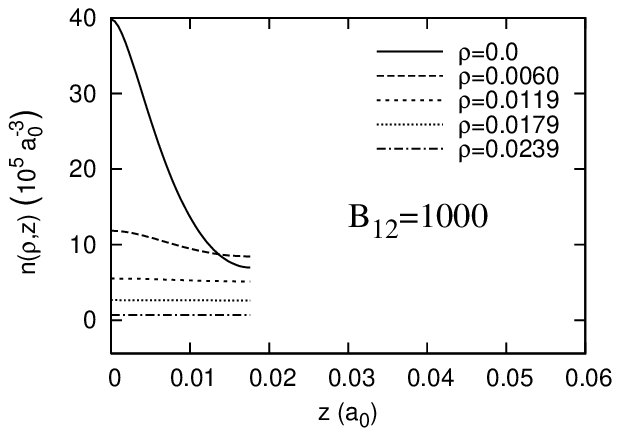}} &
\resizebox{3in}{!}{\includegraphics{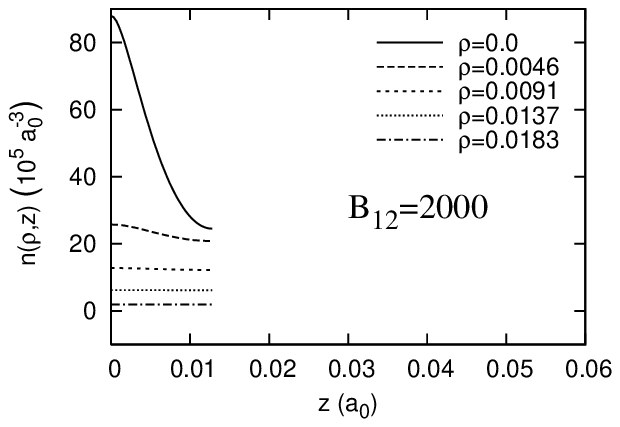}} \\
\end{tabular}
\caption{The density distribution of electrons in the iron infinite
chain at four different magnetic field strengths (labeled on the
graphs). The density is shown as a function of $z$ for five
equally spaced $\rho$ points from the center of a cell ($\rho=0$) to
the guiding center radius of the highest occupied $m$ level
($\rho=\rho_{m_{\rm max}}$). The $\rho$ points are given in units of
$a_0$.}
\label{DensityCompfig2}
\end{center}
\end{figure}

Our numerical results for Fe are given in Table~\ref{Fetable}. The
electron density profile at various field strengths is shown in
Figs.~\ref{DensityCompfig} and \ref{DensityCompfig2}. As the magnetic
field increases the density goes up, for two reasons. First, the
equilibrium ion separation decreases. Second, the electrons become
more tightly bound to each ion, in both the $\rho$ and $z$ directions
(the electrons move closer to each ion faster than the ions move
closer to each other). It is interesting to note that the peak density
at a given $z$ is not necessarily along the centeral axis of the chain
($\rho=0$), but gradually moves outward with increasing $z$.

\begin{figure}
\includegraphics[width=6.5in]{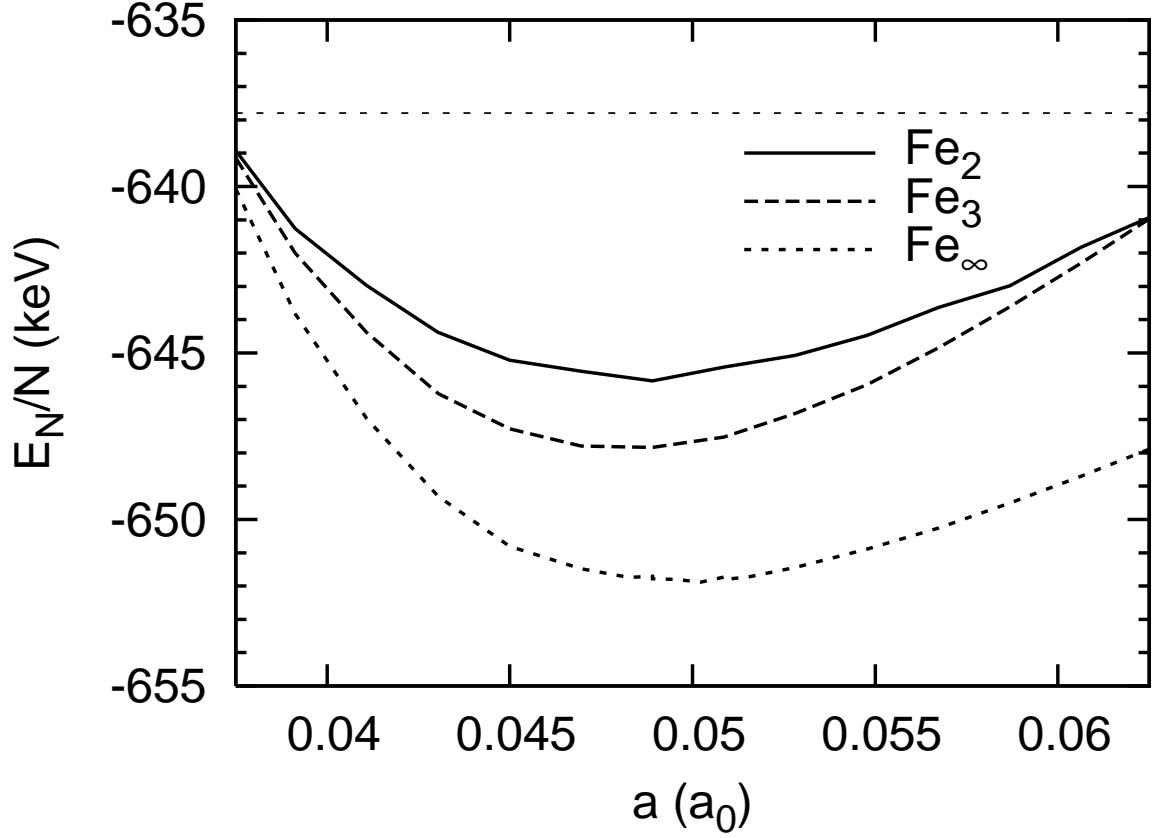}
\caption{The energy per cell as a function of the ion separation for
an infinite Fe chain at $B_{12}=500$.  The molecular energy per atom
versus ion separation for the Fe$_2$ and Fe$_3$ molecules at the same
field strength (based on calculations in paper I) are also shown. The
energy of the Fe atom is shown as a horizontal line at $-637.8$~keV.}
\label{FeChn500fig}
\end{figure}

\begin{figure}
\includegraphics[width=6.5in]{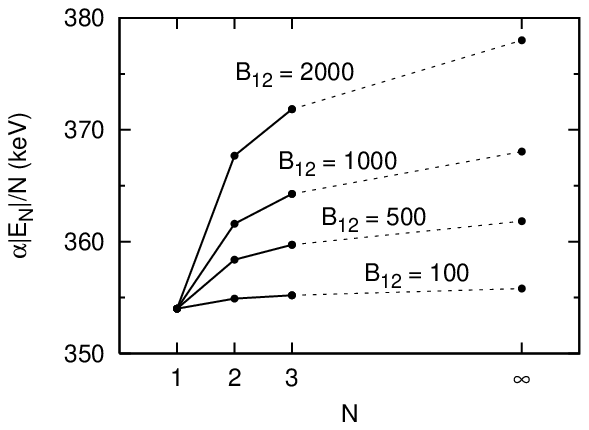}
\caption{The molecular energy per atom, $|E_N|/N$, for the Fe$_N$
molecule, as a function of $N$ at several different field strengths.
The results of finite molecules are based on paper I \citep{medin06a}.
As $N$ increases, $E_N/N$ asymptotes to $E_\infty$.  To facilitate
plotting, the values of $|E_1|$ (atom) at different magnetic field
strengths are normalized to the value at $B_{12}=100$, $354.0$ keV\@. 
This means that $\alpha = 1$ for
$B_{12}=100$, $\alpha = 354.0/637.8$ for $B_{12}=500$, $\alpha =
354.0/810.6$ for $B_{12}=1000$, and $\alpha = 354.0/1021.5$ for
$B_{12}=2000$.}
\label{FeMolfig}
\end{figure}

\begin{figure}
\includegraphics[width=6in]{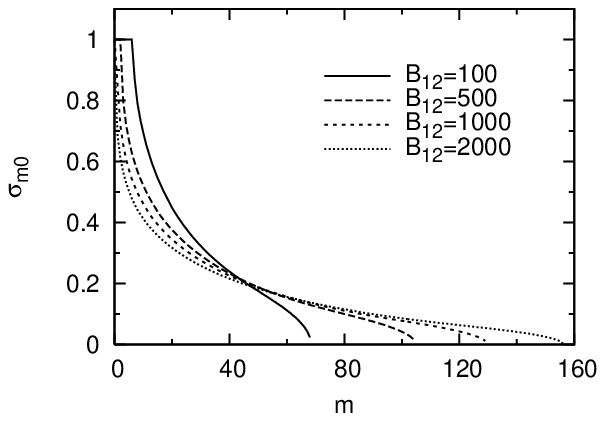}
\includegraphics[width=6in]{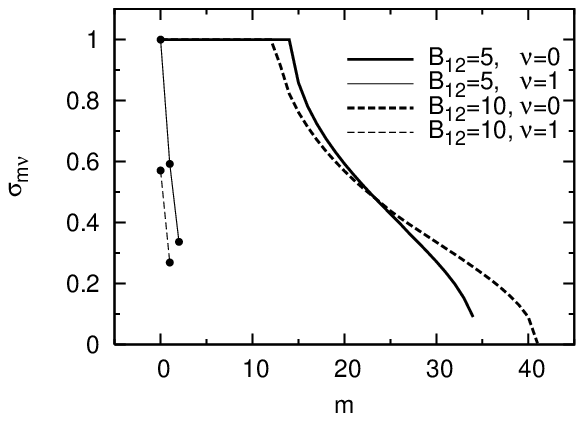}
\caption{The occupation numbers of each $m$ level of infinite Fe
chains, for various magnetic field strengths. For $B_{12}\agt 100$,
only the $\nu=0$ bands are occupied by the electrons (upper
panel). For $B_{12}=5$ and $10$ the $m$ levels with $\nu=1$ are shown
with points as well as lines, since there are only a few such occupied
levels (lower panel).}
\label{Fesigma}
\end{figure}

The energy curves for Fe$_2$, Fe$_3$ (calculated in paper I
\citep{medin06a}), and Fe$_\infty$ at $B_{12}=500$ are shown in
Fig.~\ref{FeChn500fig}. Figure~\ref{FeMolfig} compares the molecular
and infinite chain energies at various field strengths, showing that
as $N$ increases, the energy per atom in the Fe$_N$ molecule
approaches $E_\infty$ for the infinite chain. Figure~\ref{Fesigma}
gives the occupation number $\sigma_{m\nu}$ of different bands at
various field strengths. For $B_{12}\agt 100$, only the $\nu=0$ bands
are occupied; for such field strengths, the Fe atom also has all its
electrons in the tightly bound $\nu=0$ states (see
Table~\ref{Fetable}). At $B_{12}=100$, the number of fully occupied
bands is $n_f^{(0)}=7$ ($m=0,1,2,\ldots,6$, all with $\nu=0$). As $B$
increases, the electrons spread to more Landau orbitals, and the
number of occupied $m$-states $n_m^{(0)}$ increases, approximately as
$n_m^{(0)}\propto Z^{2/5}b^{1/5}$. Note that at the highest field
strength considered, the electrons occupy $m=0,1,2,\ldots,156$ ---
keeping track of all these Landau orbitals ($n_m^{(0)}=157$) is one of
the more challenging aspects of our computation. Table~\ref{Fetable}
shows that for $B_{12}\agt 100$,
\be
E_\infty \simeq -356\,B_{14}^{0.374}~{\rm keV},\qquad
a = 0.107\,B_{14}^{-0.43}~{\rm }a_0
\ee
[where $B_{14}=B/(10^{14}~{\rm G})$]. These scaling expressions are more
approximate than for H and He. The electron work function
$W=|\varepsilon_F|$ does not scale as Eq.~(\ref{fermiscale}), but is a
fraction of the ionization energy: from paper I \citep{medin06a}, the
ionization energies of Fe at $B_{12}=100,500,1000,2000$ are
$1.2$,~$2.5$,~$3.4$,~and~$5.5$~keV, respectively.

Note that at $B_{12}=5$ and $10$, the cohesive energy ($Q_\infty=E_a-
E_\infty$) of the iron chain is rather small compared to the absolute
value of the ground-state energy of the atom ($|E_a|$) or chain
($|E_\infty|$). For these field strengths, our formal numerical result
for the cohesive energy is at or smaller than the standard error of
our computations ($0.1\%$ of $|E_a|$ or $|E_\infty|$), so we have
redone the calculations using more grid and integration points such
that the atomic and chain energies reported here for these field
strengths are accurate to at least $0.02\%$ of $|E_a|$ or $|E_\infty|$
(see the Appendix). Although these more-accurate cohesive energies are
(barely) larger than the error in our calculations, there are of
course systematic errors introduced by using density functional theory
which must be considered. It is very possible that a similar,
full-band-structure calculation using Hartree-Fock theory would find
no binding. In any case, for such ``low'' field strengths ($B_{12}\alt
10$) the exact result of our one-dimensional calculation is not
crucial, since in the three-dimensional condensed matter the
additional cohesion resulting from chain-chain interactions dominates
over $Q_\infty$, as we will show in Sec.~\ref{sec:3d}.

At $B_{12}=5$, \citet{neuhauser87} and \citet{jones85} found that iron
is not bound, while we find that it is barely bound. At $B_{12}=10$,
\citet{jones86} calculated the cohesive energy for three-dimensional
condensed matter, so we compare our results with those of
Ref.~\citep{jones86} in Sec.~\ref{sec:3d}\@. We have not found any
quantitative calculations of cohesive energies for iron at field
strengths larger than $10^{13}$~G.

%%%%%%%%%%%%%
\begin{table}
\caption{The ground-state energy (per unit cell) $E_\infty$ (in units
of keV), ion separation $a$ (in units of the Bohr radius $a_0$),
electron occupation numbers ($n_m^{(0)},n_f^{(0)};n_m^{(1)},
n_f^{(1)}$), and Fermi level energy $\varepsilon_F$ (in eV) of 1D
infinite iron chains, over a range of magnetic field strengths. The
ground-state energy of individual Fe atoms, $E_a$ (in units of keV),
is also provided for reference (this is based on the
density-functional-theory calculation of Ref.~\citep{medin06a}). The
dimensionless energy $\bar E_\infty$ and ion separation $\bar a$ are
calculated using Eq.~(\ref{eq:scale}). The scaling exponents $\beta$
and $\gamma$, defined by $E_a,~E_\infty \propto B^\beta$, and
$\varepsilon_F \propto B^\gamma$, are calculated over the three
magnetic field ranges provided in the table: $B_{12}=1-10$, $10-100$,
$100-1000$ (the exponent in the $B_{12}=1$ row corresponds to the fit
over $B_{12}=1-10$, etc.). For atoms the electron configuration is
specified by the notation $(n_0,n_1)$ (with $n_0+n_1=Z=26$),
where $n_0$ is the number of electrons in the $\nu=0$ orbitals and
$n_1$ is the number of electrons in the $\nu=1$ orbitals. For infinite
chains, the occupation of different $(m\nu)$ bands is designated by
the notation $(n_m^{(0)},n_f^{(0)};n_m^{(1)},n_f^{(1)})$, where
$n_m^{(0)}$ is the total number of occupied $\nu=0$ orbitals (from
$m=0$ to $m=n_m^{(0)}-1$), and $n_m^{(1)}$ the corresponding number
for the $\nu=1$ orbitals; $n_f^{(0)}$ ($n_f^{(1)}$) is the number of
fully occupied ($\sigma_{m\nu}=1$) $\nu=0$ ($\nu=1$) orbitals. Note
that for $B_{12}\agt 100$, only the $\nu=0$ states are occupied in the
Fe atom, and only the $\nu=0$ bands are occupied in the Fe chain; see
Fig.~\ref{Fesigma}.}
\label{Fetable}
\centering
\begin{tabular}{c | r@{}l c c | r@{}l l c l c c c c}
\hline\hline
& \multicolumn{4}{| c |}{Fe} & \multicolumn{9}{| c}{Fe$_\infty$} \\
$B_{12}$ & \multicolumn{2}{| c}{$E_a$~(keV)} & $(n_0,n_1)$ & $\beta$ & \multicolumn{2}{| c}{$E_\infty$~(keV)} & \multicolumn{1}{c}{$\bar{E}_\infty$} & $\beta$ & \multicolumn{1}{c}{$a$} & \multicolumn{1}{c}{$\bar{a}$} & 
$(n_m^{(0)},n_f^{(0)};n_m^{(1)},n_f^{(1)})$ & $\varepsilon_F$~(eV) & $\gamma$ \\
\hline
5 & -107&.23 & (24,2) & 0.407 & -107&.31 & 0.522 & 0.407 & 0.42 & 4.7 & (35,15;3,1) & -161  & 0.27 \\
\hline
10 & -142&.15 & (25,1) & 0.396 & -142&.30 & 0.525 & 0.398 & 0.30 & 4.4 & (42,13;2,0) & -194  & 0.30 \\
\hline\hline
100 & -354&.0 & (26,0) & 0.366 & -355&.8 & 0.522 & 0.376 & 0.107 & 4.0 & (69,7) & -384  & 0.26 \\
\hline
500 & -637&.8 & (26,0) & 0.346 & -651&.9 & 0.503 & 0.371 & 0.050 & 3.5 & (105,2) & -583  & 0.12 \\
\hline
1000 & -810&.6 & (26,0) & 0.334 & -842&.8 & 0.493 & 0.372 & 0.035 & 3.3 & (130,1) & -635  & 0.12 \\
\hline
2000 & -1021&.5 & (26,0) & - & -1091&.0 & 0.483 & - & 0.025 & 3.1 & (157,0) & -690  & - \\
\hline\hline
\end{tabular}
\end{table}

%%%%%%%%%%%%%%%%%%%%%%%%%%%%%%%%%%%%%%%%%%%%%%%%%%%%%%%%
\section{Calculations of three-dimensional condensed matter}
\label{sec:3d}

For the magnetic field strengths considered in this paper ($B\agt
10^{12}$~G), H and He infinite chains are significantly bound relative
to individual atoms.  Additional binding energy between 3D condensed
matter and 1D chain is expected to be small \citep{lai92} (see
below). Thus the cohesive energy of the 3D condensed H or He,
$Q_s=E_a-E_s$ (where $E_s$ is the energy per cell in the 3D condensed
matter), is close to $Q_\infty=E_a-E_\infty$, the cohesive energy of
the 1D H or H chain. For C and Fe at relatively low magnetic fields
(e.g., C at $B_{12}\alt 10$ and Fe at $B_{12}\alt 100$), 1D chains are
not significantly bound relative to atoms and additional cohesion due
to chain-chain interactions is important in determining the true
cohesive energy of the 3D condensed matter. Indeed, for Fe at
$B_{12}=5,10$, our calculations of 1D chains give such a small
$Q_\infty$ (see Table~\ref{Fetable}) that it is somewhat ambiguous as
to whether the Fe condensed matter is truly bound relative to
individual atoms. In these cases, calculations of 3D condensed matter
is crucial \citep{jones86}.

In this section, we present an approximate calculation of the
relative binding energy between 3D condensed matter and 1D chains,
$\Delta E_s=E_s-E_\infty$.

%%%%%%%%%%%%%%%%%%%%%%%%%%%%%%%%%%%%%%%%%%%%
\subsection{Method}

To form 3D condensed matter we place the infinite chains in parallel bundles 
along the magnetic field. We consider a body-centered tetragonal lattice
structure; i.e., the chains are uniformly spaced in over a grid in the $xy$
plane (perpendicular to the magnetic axis), with every other chain in 
the grid shifted by half a cell ($\Delta z=a/2$) in the $z$ direction. 
The transverse separation between two nearest neighboring chains is denoted by 
$2R$, with $R$ to be determined.

To calculate the ground-state energy of this 3D condensed matter,
we assume that the electron density calculated for an individual 1D chain
is not modified by chain-chain interactions, thus we do not solve for the
full electron density in the 3D lattice self-consistently.
In reality, for each Landau orbital the transverse wave function
of an electron in the 3D lattice is no longer given by 
Eq.~(\ref{Wmeq}) (which is centered at one particular chain),
but is given by a superposition of many such Landau wave functions
centered at different lattice sites and satisfies the periodic 
(Bloch) boundary condition. The longitudinal wave function $f_{m\nu k}(z)$ 
will be similarly modified.
Our calculations show that the equilibrium separation ($2R$) between chains 
is large enough that there is little overlap in the electron densities of 
any two chains, so we believe that our approximation is reasonable. 

Using this approximation, the electron density in the 3D lattice is simply 
the sum of individual infinite chain electron densities:
\be
n_{\rm 3D}(\mathbf{r}) = \sum_{ij} n(\mathbf{r}-\mathbf{r}_{ij}) \,,
\label{3ddensityeq}
\ee
where $n(\mathbf{r})$ are the electron density in the 1D chain
(as calculated in Secs.~\ref{sec:method}-\ref{sec:result-1d}), the sum
over $ij$ spans all positive and negative integers, and
\be
\mathbf{r}_{ij} = 2Ri\,\hat{\mathbf{x}}+2Rj\,\hat{\mathbf{y}}+\frac{a}{2}[i,j]\,\hat{\mathbf{z}}
\ee
represents the location of the origin of each chain (the notation
$[i,j]=1$ when $i+j=$ odd, and $[i,j]=0$ when $i+j=$ even).
In practice, the chain-chain overlap is so small that we only need to 
consider neighboring chains. The density at a point in the positive
$xyz$ octant of a 3D unit cell is approximately given by
\be
n_{\rm 3D}(\mathbf{r}) \simeq
n(\mathbf{r})+n(\mathbf{r}-2R\hat{\mathbf{x}}-a/2\hat{\mathbf{z}})+n(\mathbf{r}-2R\hat{\mathbf{y}}-a/2\hat{\mathbf{z}})+n(\mathbf{r}-2R\hat{\mathbf{x}}-2R\hat{\mathbf{y}})\,.
\ee

The energy (per unit cell) $\Delta E_{3D}(R)$ of the 3D condensed
matter relative to the 1D chain consists of the chain-chain
interaction Coulomb energy $\Delta E_{\rm Coul}$ and the additional
electron kinetic energy $\Delta E_K$ and exchange-correlation energy
$\Delta E_{\rm exc}$ due to the (slight) overlap of different
chains. The dominant contribution to the Coulomb energy comes from the
interaction between nearest-neighboring cells.  For a given cell in
the matter, each of the eight nearest-neighboring cells contributes an
interaction energy of
\be
E_{\rm nn} = E_{eZ,\rm nn} + E_{\rm dir,nn} + E_{ZZ,\rm nn} \,,
\ee
where
\be
E_{eZ,\rm nn} = -Ze^2 \int_{|z|<a/2} d\mathbf{r} \, \frac{n(\mathbf{r})}{|\mathbf{r} - \mathbf{r}_{\rm nn}|} \,,
\label{eznneq}
\ee
\be
E_{\rm dir,nn}[n] = \frac{e^2}{2} \int \!\! \int_{|z|<a/2,\,|z'|<a/2} d\mathbf{r}\,d\mathbf{r}' \, \frac{n(\mathbf{r}) n(\mathbf{r}')}{|\mathbf{r} - (\mathbf{r}'+\mathbf{r}_{\rm nn})|}
\label{dirnneq}
\ee
\be
E_{ZZ,\rm nn} = \frac{1}{2}\frac{Z^2e^2}{|\mathbf{r}_{\rm nn}|} = \frac{1}{2}\frac{Z^2e^2}{\sqrt{(a/2)^2+(2R)^2}} \,,
\ee
and $\mathbf{r}_{\rm nn}$ is the location of the ion in a nearest-neighboring 
cell, for example
\be
\mathbf{r}_{\rm nn} = 2R\hat{\mathbf{x}}+\frac{a}{2}\,\hat{\mathbf{z}}.
\ee
More distant cells contribute to the Coulomb energy through their
quadrupole moments. The classical quadrupole-quadrupole interaction energy 
between two cells separated by a distance $d$ is 
\be
E_{QQ}(d,\theta) = \frac{3e^2}{16}
\frac{Q_{zz}^2}{d^5}(3-30\cos^2\theta+35\cos^4\theta) \,,
\ee
where $Q_{zz}$ is given by Eq.~(\ref{eq:qzz})
and $\theta$ is the angle between the line joining the two quadrupoles 
and the $z$ axis. 
The total contribution from all nonneighboring cells 
to the Coulomb energy is then 
\be 
{1\over 2}\sum_{(ijk)}E_{QQ}({\bf r}_{ijk}),
\label{eq:eqq}\ee
where 
\be
\mathbf{r}_{ijk} = \mathbf{r}_{ij}+a\,k\,\hat{\mathbf{z}}, \quad
d=|{\bf r}_{ijk}|,\quad
\cos\theta={k+[i,j]/2\over d/a},
\ee
and the sum in Eq.~(\ref{eq:eqq}) spans over all positive and negative integers
except those corresponding to the nearest neighbors.

In the density functional theory, the kinetic and exchange-correlation 
energies depend entirely on the electron density. These energies differ
in the 3D condensed matter from the 1D chain because 
the overall electron density $n_{\rm 3D}({\bf r})$ 
[see Eq.~(\ref{3ddensityeq})]
within each 3D cell is (slightly) larger than $n({\bf r})$
due to the overlap of the infinite chains. 
Since we do not solve for the electron density in the 3D condensed matter 
self-consistently,
we calculate the kinetic energy difference using the local (Thomas-Fermi)
approximation:
\be
\Delta E_K (R)= \int_{|z|<a/2;\,|x|,|y|<R} \! d\mathbf{r} \, n_{\rm 3D}(\mathbf{r})\, \varepsilon_K(n_{\rm 3D}) - \int_{|z|<a/2} \! d\mathbf{r} \, n(\mathbf{r})\, \varepsilon_K(n) \,.
\label{eq:delk}
\ee
Here $\varepsilon_K(n)$ is the (Thomas-Fermi) kinetic energy (per electron)
for an electron gas at density $n$, and is given by (e.g., \citep{lai01})
\be
\varepsilon_K(n) = 
{\hbar^2 (2\pi^2\rho_0^2n)^2\over 6m_e}=
\frac{e^2}{3\rho_0}\,b^{1/2}\, t \,,
\ee 
where $t$ is given by Eq.~(\ref{eq:tdefine}).
Note that the regions of integration in the $xy$ direction
are different for the two terms in Eq.~(\ref{eq:delk}),
as in the 1D chain the unit cell extends over all $\rho$ space, 
while in the 3D condensed matter the cell is restricted to $x,y \in [-R,R]$.

Similar to $\Delta E_K$, in the local approximation, the change in 
exchange-correlation energy per unit cell is
\be
\Delta E_{\rm exc}(R) = \int_{|z|<a/2;\,|x|,|y|<R} \! d\mathbf{r} \, n_{\rm 3D}(\mathbf{r})\, \varepsilon_{\rm exc}(n_{\rm 3D}) - \int_{|z|<a/2} \! d\mathbf{r} \, n(\mathbf{r})\, \varepsilon_{\rm exc}(n) \,,
\ee
where $\varepsilon_{\rm exc}(n)$ is the exchange-correlation energy (per electron)
at density $n$ (see Sec.~\ref{subsec:eqn}).

Combining the Coulomb energy, the kinetic energy, and the exchange-correlation 
energy, the total change in the energy per unit cell when 3D condensed 
matter is formed from 1D infinite chains can be written
\be
\Delta E_{\rm 3D}(R) = \Delta E_{\rm Coul}
+ \Delta E_K + \Delta E_{\rm exc}\,,
\label{eq:e3d}\ee
where 
\be
\Delta E_{\rm Coul}(R) = 8E_{\rm nn} + 
{1\over 2}\sum_{(ijk)}E_{QQ}({\bf r}_{ijk})\,.
\label{eq:ecoul}
\ee
We calculate $\Delta E_{\rm 3D}(R)$ as a function of $R$ and
locate the minimum to determine the equilibrium chain-chain 
separation $2R$ and the equilibrium energy of the 3D condensed matter.
Our method for evaluating various integrals is described in the appendix.

%%%%%%%%%%%%%%%%%%
\subsection{Results: 3D condensed matter}

\begin{figure}
\includegraphics[width=6.5in]{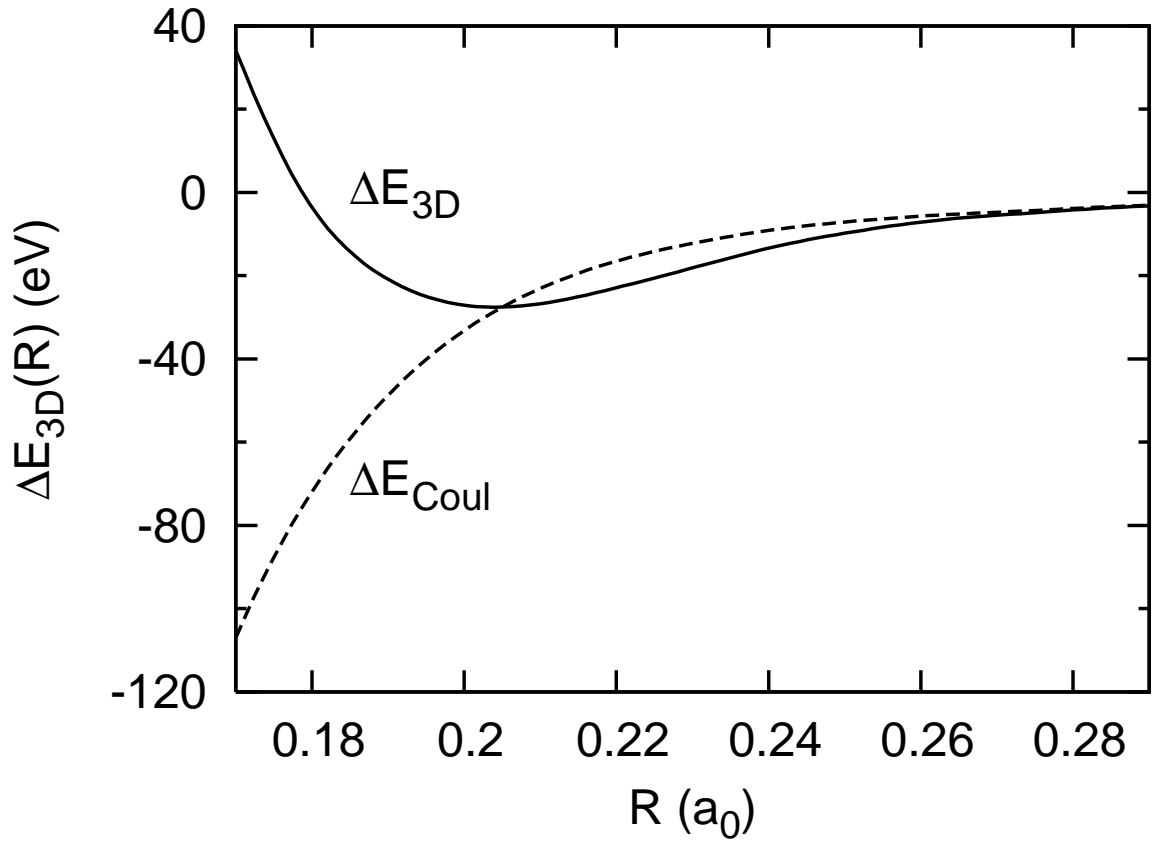}
\caption{The energy (per cell) of 3D condensed matter relative to 1D
chain as a function of $R$, for carbon at $B_{12}=1$. The
nearest-neighbor chain-chain separation in the 3D condensed matter is
$2R$. The solid curve gives $\Delta E_{\rm 3D}(R)$
[Eq.~(\ref{eq:e3d})] and the dashed curve gives only the Coulomb
energy $\Delta E_{\rm Coul}$ [Eq.~(\ref{eq:ecoul})].}
\label{C13Dfig}
\end{figure}

Table~\ref{3Dtable} presents our numerical results for the equilibrium
chain-chain separation $2R=2R_{\rm eq}$ and the energy difference (per
cell) between the 3D condensed matter and 1D chain, $\Delta
E_s=E_s-E_\infty =\Delta E_{\rm 3D}(R=R_{\rm eq})$, for C and Fe at
various magnetic field strengths.  A typical energy curve is shown in
Fig.~\ref{C13Dfig}. We see that it is important to include the
kinetic energy contribution $\Delta E_K$ to the 3D energy; without 
$\Delta E_K$, the energy curve would not have a local minimum at a finite
$R$.

A comparison of the $R$ values in Table~\ref{3Dtable} with 
various iron chain electron densities in Fig.~\ref{DensityCompfig}
shows that our assumption of small electron density 
overlap between chains is indeed a good approximation.
The electron densities are slowly-varying at the overlapping region, 
so using the local (Thomas-Fermi) model to calculate the kinetic energy 
difference is also consistent with the results of our model. Our equilibrium $R$ is within about $15\%$ of the value predicted in the uniform cylinder model [see Eq.~(\ref{chainscale})].

\begin{table}
\caption{The energy difference (per unit cell) between the 3D condensed matter
and 1D chain, $\Delta E_s=E_s-E_\infty$, for carbon and iron over a range of
magnetic field strengths. Energies are given in units of
eV for C and keV for Fe. The equilibrium chain-chain separation is
$2R$ (in units of the Bohr radius $a_0$).}
\label{3Dtable}
\centering
\begin{tabular}{c | c c | c@{}l c}
\hline\hline
 & \multicolumn{2}{c |}{C} & \multicolumn{3}{| c}{Fe} \\
$B_{12}$ & $\Delta E_s$ & $R$ & \multicolumn{2}{| c}{$\Delta E_s$} & $R$ \\
 & (eV) & & \multicolumn{2}{| c}(keV) & \\
\hline
1 & -30 & 0.200 & & & \\
\hline
5 & -40 & 0.110 & -0&.6 & 0.150 \\
\hline
10 & -20 & 0.094 & -0&.6 & 0.115 \\
\hline
100 & -20 & 0.041 & -2&.2 & 0.054 \\
\hline
500 & -30 & 0.022& -2&.1 & 0.025 \\
\hline
1000 & -10 & 0.017 & -1&.3 & 0.021 \\
\hline\hline
\end{tabular}
\end{table}

\begin{figure}
\includegraphics[width=6.5in]{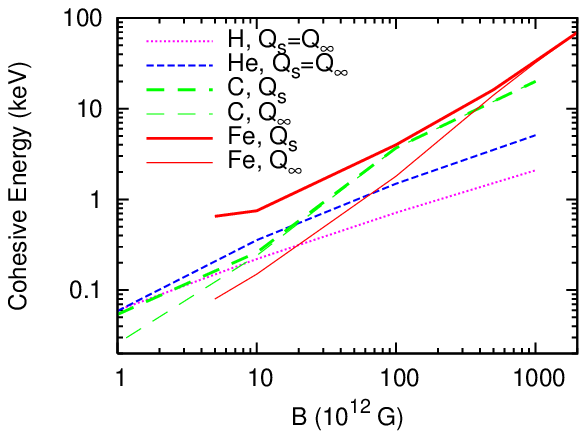}
\caption{The cohesive energy as a function of $B$, for H (dotted line) and He (short-dashed line) infinite chains and C (long-dashed lines) and Fe (solid lines) infinite chains (lighter lines) and 3D condensed matter (heavier lines).}
\label{Ecfig}
\end{figure}

Given our results for $\Delta E_s$ and the cohesive energy of 1D chains, 
$Q_\infty=E_a-E_\infty$, we can obtain the cohesive energy of
3D condensed matter from
\be
Q_s=E_a-E_s=E_a-(E_\infty+\Delta E_s)=Q_\infty-\Delta E_s \,.
\ee
For H and He, we find that $|\Delta E_s|$ is small compared to
$Q_\infty$ and thus $Q_s\simeq Q_\infty$.  Figure~\ref{Ecfig} depicts
$Q_s$ and $Q_\infty$ as a function of $B$ for H, He, C, and Fe.

The only previous quantitative calculation of 3D condensed matter is
that by \citet{jones86}, who finds cohesive energies of
$Q_s=0.60$,~$0.92$~keV for iron at $B_{12}=5,10$. At these field
strengths, our calculation (see Tables~\ref{Fetable} and
\ref{3Dtable}) gives $Q_s=E_a-E_s=Q_\infty-\Delta E_s =
0.08+0.6\simeq 0.7$~keV and $0.15+0.6\simeq 0.75$~keV, respectively.

Note that our calculations and the results presented here assume that
the ion spacing along the magnetic axis in 3D condensed matter, $a$,
is the same as in the 1D chain. We have found that if both $a$ and
$R$ are allowed to vary, the 3D condensed matter energy can be lowered
slightly. This correction is most important for relatively low
field strengths. For example, in the case of Fe at $B_{12}=10$, if we
increase $a$ from the 1D chain value by $10\%$, then $Q_\infty$
decreases by about $50$~eV, but $|\Delta E_s|$ increases by about
$200$~eV, so that $Q_s$ is increased to $\sim 0.9$~keV\@. Given the
approximate nature of our 3D calculations, we do not explore such
refinement in detail in this paper.

%%%%%%%%%%%%%%%%%%%%%%%%%%%%%%%%%%%%%%%
\section{Discussions}

Using density functional theory, we have carried out extensive
calculations of the cohesive properties of 1D infinite chains and 3D
zero-pressure condensed matter in strong magnetic fields. Our results,
presented in various tables, figures, and fitting formulae, show that
hydrogen, helium, and carbon infinite chains are all bound relative to
individual atoms for magnetic fields $B\ge 10^{12}$~G, but iron chains
are not (significantly) bound until around $B\sim 10^{14}$~G\@. For a
given zero-pressure condensed matter system, the cohesion along the
magnetic axis (chain axis) dominates over chain-chain interactions
across the magnetic axis at sufficiently strong magnetic fields. But
for relative low field strengths (e.g. Fe at $B\alt 10^{14}$~G and C
at $B\alt {\rm a~few}\times 10^{12}$~G), chain-chain interactions play
an important role in the cohesion of 3D condensed matter. Our
calculations show that for the field strengths considered in this
paper ($B\agt 10^{12}$~G), 3D condensed H, He, C and Fe are all bound
relative to individual atoms: For C, the cohesive energy $Q_s=E_a-E_c$
ranges from $\sim 50$~eV at $B=10^{12}$~G to $20$~keV at $10^{15}$~G;
for Fe, $Q_s$ ranges from $\sim 0.8$~keV at $10^{13}$~G to $33$~keV at
$10^{15}$~G.

Our result for the 1D infinite chain energy (per cell), $E_\infty$, is
consistent with the energies of finite molecules obtained in paper I
\citep{medin06a}, where we showed that the binding energy (per atom)
of the molecule, $|E_N|/N$ (where $E_N$ is the ground-state energy and
$N$ is the number of atoms in the molecule), increases with increasing
$N$, and asymptotes to a constant value. The values of $|E_N|/N$ for
various molecules obtained in Ref.~\citep{medin06a} are always less
than $|E_\infty|$. Since the electron energy levels in a finite
molecule and those in an infinite chain are quite different (the
former has discrete states while the latter has band structure), and
the computations involved are also different, the consistency between
the finite molecule results and 1D chain results provides an important
check for the validity of our calculations.

It is not straightforward to assess the accuracy of our
density-functional-theory calculations of infinite chains compared to
the Hartree-Fock method. For finite molecules with small number of
electrons, using the available Hartree-Fock results, we have found
that density functional theory tends to overestimate the binding
energy by about $10\%$, although this does not translate into an
appreciable error in the molecular dissociation energy
\citep{medin06a}. For infinite chains, the only previous calculation
using the Hartree-Fock method \citep{neuhauser87} adopted an
approximate treatment for the electron band structure (e.g., assuming
that the electron energy increases as $k^2/2$ as the Bloch wave number
$k$ increases), which, as we showed in this paper
(Sec.~\ref{subsec:complex}), likely resulted in appreciable error to
the total chain energy. Since the cohesive energy $Q_\infty$ of the
chain involves the difference in the binding energy the 1D chain and
the atom, and because of the statistical nature of density functional
theory, we expect that our result for $Q_\infty$ is more accurate for
heavy elements (C and Fe) than for light elements (H and He). We note
that it is very difficult (perhaps impractical) to carry out {\it ab
initio} Hartree-Fock calculations of infinite chains if no
approximation is made about the electron band structure. This is
especially the case in the superstrong magnetic field regime where
many Landau orbitals are populated. For example, for the Fe chain at
$B=10^{15}$~G, one must be dealing with $130$ Landau orbitals (see
Table~\ref{Fetable}), each with its own band structure --- this would
be a formidable task for any Hartree-Fock calculation.

We also note that our conclusion about 3D condensed matter is not
based on fully self-consistent calculations and uses several
approximations (Sec.~\ref{sec:3d}). Although we have argued that the
approximations we adopted are valid and our calculation gave
reasonable values for the relative binding energies between 1D chains
and 3D condensed matter, it would be desirable to carry out more
definitive calculations of 3D condensed matter.

Our computed binding energies and equilibrium ion separations of
infinite chains and condensed matter agree approximately with the
simple scaling relations (e.g., $E_\infty$ and $a$ as a function of
$B$) derived from the uniform gas model (Sec.~\ref{sec:basic}). We have
provided more accurate fitting formulae which will allow one to obtain
the cohesive energy at various field strengths. Our result for the
electron work function ($W=|\varepsilon_F|$), however, does not agree
with the simple scaling relation derived for the uniform electron gas
model. For example, we found that $W$ scales more slowly with $B$
($\gamma$ is significantly smaller than than $2/5$) and does not
depend strongly on $Z$ (as opposed to the $Z^{4/5}$ dependence for the
uniform gas model); see Tables~\ref{Htable}--\ref{Fetable}. This
``discrepancy'' is understandable since, unlike the $B=0$ case, in
strong magnetic fields the ionization of an atom and binding energy of
condensed matter can be very different in values and have different
dependences on $B$: for sufficiently large $B$, the former scales
roughly as $(\ln b)$, while the later scales as $\sim b^{0.4}$. Our
computed electron work function is of order (and usually a fraction
of) the ionization energy of the corresponding atom, which is
generally much smaller than the estimate of $W$ based on uniform gas
model. We also found that the ionization energy of successively larger
(finite) molecules \citep{medin06a} approaches our calculated work
function for the infinite chain --- thus we believe our result for $W$
is reliable.  Note that \citet{jones86} also found that the work
function $W$ is almost independent of $Z$, but his $W$ values scale as
$B^{0.5}$ and are much larger than our results for the same field
strengths. His $W$ values are also larger than the ionization
energies of the corresponding atoms.

Our results for the cohesive energy and work function of condensed
matter in strong magnetic fields have significant implications for the
physical conditions of the outermost layers of magnetized neutron
stars and the possible existence of ``vacuum gap'' accelerators in
pulsars. We plan to investigate these issues in the future.

\begin{acknowledgments}
We thank Neil Ashcroft for useful discussion. This work has been
supported in part by NSF Grant No. AST 0307252, NASA Grant No. NAG 5-12034 and
{\it Chandra} Grant No. TM6-7004X (Smithsonian Astrophysical Observatory).
\end{acknowledgments}

%%%%%%%%%%%%%%%%%%%%%%%%%%%%%%%%%%%%%%%%%%%%%%%%%%%%%%%%%%%%%%%%%%
\appendix

\section{Technical details and numerical method}

\subsection{Evaluating the integrals in the Kohn-Sham equations}

The most computation-intensive term in the modified Kohn-Sham equations [Eqs.~(\ref{kohneq}) and (\ref{modkohneq2})] is the direct electron-electron interaction term
\be
V_{ee,m}(z) = \int \!\! \int_{|z'|<a(N_Q+1/2)}
d\mathbf{r}_\perp\,d\mathbf{r}' \,
\frac{|W_m|^2(\rho)\, n(\mathbf{r}')}{|\mathbf{r} - \mathbf{r}'|} \,.
\label{Veemeq}
\ee
The evaluation of this term is the rate-limiting step in the entire energy calculation. The integral is over four variables ($\rho$, $\rho'$, $z'$, and $\phi$ or $\phi-\phi'$), so it requires some simplification to become tractable. To simplify the integral we use the identity (e.g., Ref.~\citep{jackson98})
\be
\frac{1}{|\mathbf{r} - \mathbf{r}'|} =
\sum_{n=-\infty}^\infty \int_0^\infty dq \,
e^{in(\phi-\phi')} J_n(q\rho) J_n(q\rho') e^{-q|z-z'|} \,,
\label{1overreq}
\ee
where $J_n(z)$ is the $n$th order Bessel function of the first kind. Then
\ba
V_{ee}(\mathbf{r}) & = & \int_{|z'|<a(N_Q+1/2)} d\mathbf{r}' \,
\frac{n(\mathbf{r}')}{|\mathbf{r} - \mathbf{r}'|} \\
 & = & 2\pi \int_{-a(N_Q+1/2)}^{a(N_Q+1/2)} dz' \int_0^\infty dq \,
J_0(q\rho) \left[\int_0^\infty \rho'\,d\rho' \, n(\rho',z') J_0(q\rho')\right]
\exp(-q|z-z'|) \,,
\ea
and
\ba
V_{ee,m}(z) & = & \int d\mathbf{r}_\perp \,
|W_m|^2(\rho)\, V_{ee}(\mathbf{r}) \\
 & = & 4\pi^2 \int_{-a(N_Q+1/2)}^{a(N_Q+1/2)} dz' \int_0^\infty dq \,
\left[\int_0^\infty \rho\,d\rho \, |W_m|^2(\rho)\, J_0(q\rho)\right]
\left[\int_0^\infty \rho'\,d\rho' \, n(\rho',z') J_0(q\rho')\right]
\exp(-q|z-z'|) \,. \nonumber\\
\label{Veereq}
\ea
Using Eq.~(\ref{densityeq}) for the electron density distribution,
Eq.~(\ref{Veereq}) becomes
\be
V_{ee,m}(z) = \sum_{m'\nu'} \int_{-a(N_Q+1/2)}^{a(N_Q+1/2)} dz' \,
\bar{f}_{m'\nu'}^{\,2}(z') \int_0^\infty dq \, G_m(q) G_{m'}(q)
\exp(-q|z-z'|) \,,
\label{Veezeq}
\ee
where
\ba
G_m(q) & = & 2\pi \int_0^\infty \rho\,d\rho \, |W_m|^2(\rho)\, J_0(q\rho) \\
 & = & \exp(-q^2/2) L_m(q^2/2) \,,
\ea
and
\be
L_m(x) = \frac{e^x}{m!}\frac{d^m}{dx^m}(x^m e^{-x})
\ee
is the Laguerre polynomial of order $m$. These polynomials can be
calculated using the recurrence relation
\be
m L_m(x) = (2m-1-x) L_{m-1}(x) - (m-1) L_{m-2}(x) \,,
\ee
with $L_0(x)=1$ and $L_1(x)=1-x$.

Using the method outlined above the original four-dimensional integral
in Eq.~(\ref{Veemeq}) reduces to a two-dimensional integral. Once a
value for $z$ is specified, the integral can be evaluated using a
quadrature algorithm (such as the Romberg integration method
\citep{jackson98}).

\subsection{Evaluating the integrals in the calculation of 3D condensed matter}

For the 3D condensed matter calculation, we simplify the energy integrals of the nearest-neighbor interactions in a way similar to that for the infinite chain calculation. To do this, we require Eq.~(\ref{1overreq}) and one additional identity of Bessel functions:
\be
J_0(q\sqrt{a^2+b^2-2ab \cos{\theta}}) = \sum_{n=-\infty}^\infty e^{in\theta} J_n(qa) J_n(qb) \,.
\ee
With these equations the ion-electron nearest-neighbor energy term [Eq.~(\ref{eznneq})] becomes
\ba
E_{eZ,\rm nn}[n] & = & -Ze^2 \int_{|z|<a/2} d\mathbf{r} \, \frac{n(\mathbf{r})}{|\mathbf{r} - \mathbf{r}_{\rm nn}|} \\
 & = & -Ze^2 2\pi \int_{-a/2}^{a/2} dz \int_0^\infty dq \, J_0(2Rq) \left[\int_0^\infty \rho\,d\rho \, n(\rho,z) J_0(q\rho)\right] \exp(-q|z-a/2|) \\
 & = & -Ze^2 \sum_{m\nu} \int_{-a/2}^{a/2} dz \, \bar{f}_{m\nu}^{\,2}(z) \int_0^\infty dq \, J_0(2Rq) G_m(q) \exp(-q|z-a/2|) \,.
\ea
The electron-electron energy term [Eq.~(\ref{dirnneq})] becomes
\ba
E_{\rm dir,nn}[n] & = & \frac{e^2}{2} \int \!\! \int_{|z|<a/2,\,|z'|<a/2} d\mathbf{r}\,d\mathbf{r}' \, \frac{n(\mathbf{r}) n(\mathbf{r}')}{|\mathbf{r} - (\mathbf{r}'+\mathbf{r}_{\rm nn})|} \\
 & = & \frac{e^2}{2} 2\pi \int_{-a/2}^{a/2} dz \int_{-a/2}^{a/2} dz' \int_0^\infty \rho'\,d\rho' \, n(\rho',z') \times \nonumber\\
 & & \qquad \int_0^{2\pi} d\phi' \int_0^\infty dq \, J_0(q|\mathbf{r}'_\perp+\mathbf{r}_{\perp,\rm nn}|) \left[\int_0^\infty \rho\,d\rho \, n(\rho,z) J_0(q\rho)\right] e^{-q|z-z'-a/2|} \,, \nonumber\\
\ea
where $\theta$ is the angle of $\mathbf{r}'_\perp+\mathbf{r}_{\perp,\rm nn}$ in the $(\rho,\phi,z)$ cylindrical coordinate system $\Rightarrow$
\ba
E_{\rm dir,nn}[n] & = & \frac{e^2}{2} 4\pi^2 \int_{-a/2}^{a/2} dz \int_{-a/2}^{a/2} dz' \times \nonumber\\
 & & \qquad \int_0^\infty dq \, J_0(2Rq) \left[\int_0^\infty \rho\,d\rho \, n(\rho,z) J_0(q\rho)\right] \left[\int_0^\infty \rho'\,d\rho' \, n(\rho',z') J_0(q\rho')\right] e^{-q|z-z'-a/2|} \\
 & = & \frac{e^2}{2} \sum_{m\nu,m'\nu'} \int_{-a/2}^{a/2} dz \, \bar{f}_{m\nu}^{\,2}(z) \int_{-a/2}^{a/2} dz' \, \bar{f}_{m'\nu'}^{\,2}(z') \int_0^\infty dq \, J_0(2Rq) G_m(q) G_m'(q) \exp(-q|z-z'-a/2|) \,. \nonumber\\
\ea
Notice that the infinite chain expression for the nearest-neighbor electron-electron interaction energy is recovered when $R=0$ and $a/2$ is replaced by $\pm a$.

\subsection{Solving the differential equations and the total energy 
self-consistently}

The Kohn-Sham equations [Eqs.~(\ref{kohneq}) and (\ref{modkohneq2})]
are solved on a grid in $z$. Because of symmetry we only need to
consider $z\ge0$, with $z=0$ coincident with an ion. The number and
spacing of the $z$ grid points determine how accurately the equations
can be solved. In this paper we have attempted to calculate
ground-state chain energies to better than $0.1\%$ numerical
accuracy. This requires approximately (depending on $Z$ and $B$) 33
grid points for each unit cell and 3 cells (for $N_Q=1$ there are
three cells that require exact treatment: the cell under consideration
$z \in [-a/2,a/2]$ and its nearest neighbors; the rest of the cells
enter the calculation only through their quadrupole moments).  The
grid spacing is chosen to be constant from the center out to the edge
of the cell. The shape of the wave function is found within one cell
and then copied to the other cells.

For integration with respect to $\rho$, $\rho'$, or $q$ (e.g., when
calculating the direct electron-electron interaction term), our goal
of $0.1\%$ accuracy for the total energy requires an accuracy of
approximately $10^{-5}$ in the integral. A variable-step-size
integration routine is used for each such integral, where the number
of points in the integration grid is increased until the error in the
integration is within the desired accuracy.

We discussed the boundary conditions for the wave function solutions to
the Kohn-Sham equations (see Sec.~\ref{subsec:complex}). The only other
requirement we have for these wave functions is that the magnitude of
each wave function has the correct number ($\nu$) of nodes per cell
(see Fig.~\ref{wshapes}). In practice, to find $f_{m\nu 0}(z)$ and
$f_{m\nu \pi/a}(z)$ we integrate Eqs.~(\ref{kohneq2}) and
(\ref{modkohneq2}) from one edge of the $z$ grid (e.g., $z=a/2$) and
shoot toward the center ($z=0$), adjusting
$\varepsilon_{m\nu}(k=0)$ and $\varepsilon_{m\nu}(\pi/a)$ until the
correct boundary condition is satisfied. For the other $k$ values with
energies between these two extremes, we use the given energy to find a
wave function and calculate the $k$ that solves the boundary condition
Eq.~(\ref{bceq}), as discussed in Sec.~\ref{subsec:complex}.

There are two parts to our procedure for finding $f_{m\nu k}(z)$,
$\varepsilon_{m\nu}(k)$, and $\sigma_{m\nu}$ self-consistently: (i)
determining the longitudinal wave functions $f_{m\nu k}$ and periodic
potential self-consistently, and (ii) determining the electron level
occupations $\sigma_{m\nu}$ self-consistently.

To determine the $f_{m\nu k}$ wave functions self-consistently, a trial
set of wave functions and $\sigma_{m\nu}$ values is first used to
calculate the potential as a function of $z$, and that potential is
used to calculate a new set of wave functions. These new wave functions
are then used to find a new potential, and the process is repeated
until consistency is reached. In practice, we find that $f_{m\nu
k}(z)=0$ works well as the trial wave function and a linear spread of
$\sigma_{m\nu}$ from $\sigma_{0\nu}=1$ to $\sigma_{n_m \nu}=0$ works
well for the trial $\sigma$ values. Convergence can be achieved in four
or five iterations. To prevent overcorrection from one iteration to the
next, the actual potential used for each iteration is a combination of
the newly-generated potential and the old potential from the previous
iteration (the weighting used is roughly $30\%$ old, $70\%$ new).

To determine the $\sigma_{m\nu}$ level occupations self-consistently,
we first find the wave functions and eigenvalues
$\varepsilon_{m\nu}(k)$ as a function of $k$ self-consistently as
described above. With this information, and given a Fermi level energy
$\varepsilon_F$, we can calculate new $\sigma$ values, using the
equations in Sec.~III\@. The Fermi level energy is adjusted until $\sum
\sigma_{m\nu} =Z$ using Newton's method. These new $\sigma_{m\nu}$
values are used to re-calculate the wave functions
self-consistently. This process is repeated until self-consistency is
reached, which is typically after about three (for hydrogen at
$10^{12}$~G) to twelve (for iron at $2\times10^{15}$~G) full
iterations.

\end{document}